\documentclass{article}

    \PassOptionsToPackage{numbers, compress}{natbib}

\usepackage[final]{neurips_2022}

\usepackage[utf8]{inputenc} 
\usepackage[T1]{fontenc}    
\usepackage{hyperref}       
\usepackage{url}            
\usepackage{booktabs}       
\usepackage{nicefrac}       
\usepackage{microtype}      
\usepackage{xcolor}         

\usepackage{amsmath,amsfonts}
\usepackage{amsthm}
\usepackage{bbding}
\usepackage{graphicx}
\usepackage{textcomp}
\usepackage{makecell}
\usepackage{overpic}
\usepackage{subfigure}
\usepackage{amssymb}
\usepackage[utf8]{inputenc}
\usepackage[ruled,vlined]{algorithm2e}
\newtheorem{example}{Example}

\newtheorem{proposal}{Proposal}
\newtheorem{definition}{Definition}
\newtheorem{problem}{Problem}
\newcommand{\tabincell}[2]{\begin{tabular}{@{}#1@{}}#2\end{tabular}}

\title{AnomMAN: Detect Anomaly on Multi-view Attributed Networks}

\author{
  Ling-Hao Chen\thanks{Ling-Hao Chen: \texttt{thu.lhchen@gmail.com}}\\
  Tsinghua University, Xidian University\\
\And
  He Li\\
  Xidian University\\
\And
  Wanyuan Zhang\\
  Xidian University\\
\AND
  Jianbin Huang\\
  Xidian University\\
\And
  Xiaoke Ma\\
  Xidian University\\
\And
  Jiangtao Cui\\
  Xidian University\\
\AND
  Ning Li\\
  Xi'an University of Technology\\
\And
  Jaesoo Yoo\\
  Chungbuk National University\\
}

\begin{document}

\maketitle

\begin{abstract}
Anomaly detection on attributed networks is widely used in online shopping, financial transactions, communication networks, and so on. However, most existing works trying to detect anomalies on attributed networks only consider a single kind of interaction, so they cannot deal with various kinds of interactions on multi-view attributed networks. It remains a challenging task to jointly consider all different kinds of interactions and detect anomalous instances on multi-view attributed networks. In this paper, we propose a graph convolution-based framework, named  \textbf{AnomMAN}, to detect \textbf{Anom}aly on \textbf{M}ulti-view \textbf{A}ttributed \textbf{N}etworks. To jointly consider attributes and all kinds of interactions on multi-view attributed networks, we use the attention mechanism to define the importance of all views in networks. Since the low-pass characteristic of graph convolution operation filters out most high-frequency signals (aonmaly signals), it cannot be directly applied to anomaly detection tasks. AnomMAN introduces the graph auto-encoder module to turn the disadvantage of low-pass features into an advantage. According to experiments on real-world datasets, AnomMAN outperforms the state-of-the-art models and two variants of our proposed model. 
\end{abstract}

\section{Introduction}

Anomaly detection {\cite{li2020copula}} is an important task in the data mining area, and is widely applied in web shopping \cite{li2019trust}, financial transaction {\cite{anjaneyulu2019financial}}, traffic {\cite{li2021detectornet, li2020graphsanet, li2022deep}} and communication {\cite{shang2014modbus}} networks. With the emergence of a large scale of network interaction data, many researchers {\cite{Eswaran2018spotlight, ding2019deep}} tried to detect anomaly instances on social networks in recent years. However, most of these algorithms cannot deal with complex interactions {on} multi-view attributed networks~\cite{zhu2022bars, fan2020one2multi}. In fact, there are multiple kinds of interactions in the real world, such as {making comments}, purchasing goods, and adding goods into the shopping cart in online shopping networks. Existing models are not specifically designed for various kinds of schemas in networks with multiple  interactions, which makes it difficult to capture the relationship {among} various interactions.

\begin{figure}
    \centering
    \includegraphics[scale=0.78]{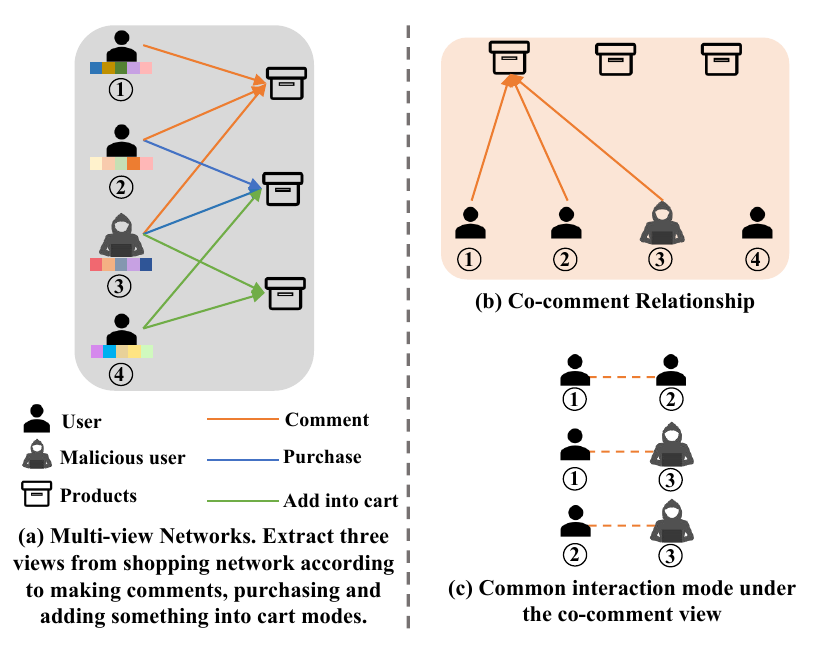}
    \caption{A toy example of online shopping multi-view attributed networks.}
    \label{fig:toy_example}
\end{figure}

Multi-view attributed networks are widely used in the real world. {On} multi-view attributed networks, {different types of interactions correspond to different methods of anomaly detection}. For instance, as shown in Figure \ref{fig:toy_example}, {user} attributes and interactions in online shopping networks can be abstracted as multi-view attributed networks. In this toy example, there are three interaction actions: purchasing goods, making comments on goods, and adding goods to his/her cart. All these actions constitute different views {on} multi-view attributed networks. A large variety of interactions {on} multi-view attributed networks bring more perspectives to analyze anomalous behaviors of users in networks.

In an anomaly detection task, anomalous instances often post positive comments or click farming on one or some specific commodities and benefit from this kind of anomalous action. Making comments and clicking actions are both anomalous actions different from normal behaviors. We can obtain multi-view web shopping networks according to different kinds of actions. As can be seen, different interactions from different views can bring us rich semantics to detect anomalous instances. Therefore, it is very important to take various kinds of interaction actions into consideration in such anomaly detection tasks.

{On} multi-view attributed networks, anomalous users may have different actions in different views. Most existing models can only detect anomalies on attributed networks with a single view. However, multi-view attributed networks can be more complex, because some instances may behave more anomalously in a specific view and less anomalously in other views. Therefore, we need to define the importance of different views to mine more anomalous actions from attributed networks. If only a single view is used to detect anomalous instances, the semantics of interaction actions will be lost.

Recently, as a deep non-linear model, Graph Neural Networks (GNNs) \cite{zhou2018graph, chen2022learning} show strong performance on many data mining tasks, such as nodes classification \cite{kipf2016semi} and clustering {\cite{bo2020structural}}. However, most algorithms based on GNNs are not specially designed for anomaly detection tasks, they simply treat the anomaly detection task as a two-class classification task. {On} real-world datasets, anomalous instances account for only a small percentage and most of them lack ground truth labels in the training stage. This phenomenon causes most GNN-based algorithms to fail to treat anomaly detection as such a simple classification task. Therefore, in this work, we propose an unsupervised model which can detect anomalies on multi-view attributed networks without the ground truth {on} the dataset.

To take both attributes and network structures into consideration, we propose a graph convolution-based framework to detect anomalies instances. However, the graph convolution operation \cite{kipf2016semi} is a low-pass filter, which filters most anomalous (high-frequency) signals {\cite{ren2019time}}. Since high-frequency signals are obviously filtered by directly using a multi-layer Graph Convolutional Network (GCN), it is challenging to detect anomalous nodes in the network. However, we can take advantage of its low-pass filter property by designing an auto-encoder-based anomaly detection model. After filtering most high-frequency signals, the more anomalous the nodes are, the {larger} their reconstruction error is. In such a manner, we transform this property of graph convolution into an advantage of our model.
To be more specific, we design a graph convolution-based auto-encoder to detect anomalies and compare the distance between the input and output of the auto-encoder to gain reconstruction error. Instances with a larger error are considered to be more anomalous ones. Based on this, we leverage a simpler but more powerful low-pass filter to replace the traditional multi-layer GCN as the encoder part of our model.

In this paper, we design a framework to detect anomalies on multi-view attributed networks. However, it is a {demanding} task to accomplish because it faces three main challenges. (1): How to capture anomalous actions from different views and mine the relationships among them? Multi-view attributed networks imply abundant interaction actions and semantics. It is not easy to extract patterns from different views into the same feature space and fuse them together. (2): In order to consider the attributes and the network structure together to detect anomalies, we introduce GCN into our model. However, it is not specifically designed for anomaly detection tasks because it {simply} treats this as a two-class classification task. (3): As a low-pass filter, GCN filters out high-frequency signals of anomalous instances, which brings difficulty to anomaly detection tasks. However, we try to turn this disadvantage into an advantage of our model.

In this paper, we propose a graph convolution-based framework, AnomMAN (Detect \textbf{Anom}aly on \textbf{M}ulti-view \textbf{A}ttributed \textbf{N}etworks), to detect anomalies over multi-view attributed networks. Our contributions can be summarized as follows:

\begin{itemize}

    \item {
         {We propose a model which} can learn the different distributions of data from different views and fuse the latent representations from different views to mine the relationships {among} them. To the best of our knowledge, it is the first time to detect anomalies on multi-view attributed networks that come from a bipartite graph.
    }

    \item{
        We propose an unsupervised model which can detect anomalies without the ground truth to accomplish the unbalanced number of labels.
    }

    \item{
        We take advantage of the “\textit{shortcomings}” of graph convolution as a low-pass filter in this anomaly detection task, and use a simpler but more powerful low-pass filter to detect anomalies.}

    \item{
        We evaluate our model on real-world datasets with both structure and attribute anomalies. Experiments show the superiority of our model in anomaly detection tasks on multi-view attributed networks.
    }
\end{itemize}

The rest of this paper is organized as follows. Section \ref{relatedwork} introduces the related work of this work. Section \ref{Preliminaries} defines the problem definition of this paper. In Section \ref{Method}, we describe the proposed model and algorithm in detail. Section \ref{experiment} reports the experimental results and some detailed discussions. We conclude our work in Section \ref{conclusion}.

\section{Related Work \label{relatedwork}}

Our work is related to anomaly detection on plain networks, anomaly detection on attributed networks, multi-view learning, and Graph Neural Networks.

\subsection{Anomaly Detection on Plain Networks}

Plain networks are modeled to represent the structural information of real-world networks. In the early years, leading scholars in the area usually did not utilize deep learning-based methods to detect anomalies in networks. These researches mainly detect anomalies by calculating the statistics of each node in the networks based on some prior assumptions. Most of these methods try to simplify the problem of anomaly detection on plain networks to the traditional anomaly detection problem which has been deeply studied. {\cite{hooi2016fraudar}} {utilizes} the outdegree or indegree indicators of the graph structure as features to detect the anomalies of network structure. With the development of graph representation learning {\cite{wang2017community}}, some researches \cite{bandyopadhyay2019outlier, bandyopadhyay2020outlier} utilize the learned embedding as the features of nodes to detect anomalies and have been used for anomaly detection performance validation. However, relying only on the structural features of networks is not enough. Therefore, we also take the problem of anomaly detection on attributed networks as our related work.

\subsection{Anomaly Detection on Attributed Networks}

{Each node in attributed networks has a feature vector which means the node attribute based on plain networks. There are many researches about embedding techniques on attributed networks. NEEC~\cite{huang2018exploring} introduces hard-to-model but valuable expert cognition into network embedding by asking experts questions.} MTSN \cite{liu2021motif} utilizes a temporal shift mechanism with GNNs to obtain dynamic and motif-based embedding. {BiANE \cite{huang2020biane} studies embedding on bipartite attributed networks by learning the inter-partition proximity and the intra-partition proximity simultaneously. Appropriate embedding makes the downstream tasks such as anomaly detection efficient \cite{xia2023moderate}.}
However, {most researches directly focused on anomaly detection on attributed networks.} For example, AMEN \cite{perozzi2016scalable} considers attributes of nodes in the {networks}, and detects anomalies of neighbor nodes. There are also some methods {which} adopt residual analysis to detect anomalies on attributed networks. Radar \cite{li2017radar} and ANOMALOUS \cite{peng2018anomalous} characterize the residual of the attribute information and its consistency with network information to detect anomalies. After that, Dominant \cite{ding2019deep} uses graph auto-encoder to detect anomalies considering the problems of network sparsity and data non-linearity. Li \textit{et al.}~\cite{li2019specae} {proposed} SpecAE with a deconvolution decoder to reconstruct and detect anomalies by density estimation. CoLA \cite{liu2021anomaly} adopts a contrastive learning-based algorithm to detect anomalies with local subgraphs coming from random walks with restart (RWR). Differing from the above, HO-GAT \cite{huang2021hybrid} detects both node-level and subgraph-level anomalies. HO-GAT designs a specialized autoencoder to simultaneously learn representations about the nodes and motif instances. {GAE \cite{du2022graph} changes the distribution of data according to the similarities among nodes, which leads abnormal nodes to deviate from normal ones.} However, the models mentioned above cannot detect anomalies in multi-view networks {scenarios}.



\subsection{Multi-view Learning on Networks}
Despite {these algorithms have made progress, they are unable to} detect anomalies on multi-view attributed networks or capture users’ different anomalous interactive actions. Multi-view attributed networks are more suitable for the real scenario since there are more than one perspective in real-world social networks.

{Multi-view learning is applied in many scenarios, and the multi-view network is one of them. MVRL \cite{WANG2021700} utilizes matrix factorization of the weighted combination of multiple views. SMGCN \cite{WU2022142} integrates multi-view graph representation and graph convolutional representation learning into a whole.}
The research on multi-view {anomaly detection} in networks can be divided into two categories. One is {to build} different views by sub-sample methods, such as SL-GAD \cite{zheng2021generative} and Sub-CR \cite{zhang2022reconstruction}. These methods utilize contextual information and sample two sub-graphs for each node to construct different views. SL-GAD samples two sub-graphs from the original graph by RWR and Sub-CR samples from a global view by leveraging graph diffusion algorithms. After that, both models maximize the agreement between node representations and their contexts by graph contrastive learning. 
The other is to build different views by dividing {attributes of nodes} collected from different feature extractors into different views. A node will be detected as an anomalous instance in some views if it deviates from the majority in these views and it will be recognized as a normal node in other views. Since the users’ attributes are different {in} different views, it’s necessary to model the whole network from a multi-view perspective. In such a manner, ALARM \cite{peng2020deep} proposes a multi-view learning-based model in the homogeneous graph, where only attributes differ in each view. We propose the AmonMAN for the multi-networks with different structures which constructs views according to users' interactions and leads to distinct structures and identical attributes in each view.

\subsection{Graph Neural Networks}

With the great progress achieved in deep learning \cite{xia2022pluralistic}, Graph Neural Networks (GNNs) are applied to Graph Representation Learning {\cite{chen2020graph}} area, aiming to deal with arbitrary graph-structured data introduced in \cite{scarselli2008graph}. Different from the traditional graph embedding model {\cite{ou2016asymmetric}}, GNNs combine attributes and network structures together to accomplish downstream tasks. Spectral Graph CNN \cite{bruna2013spectral} proposes the convolution operation on the graph based on the spectrum of the Graph Laplacian. In order to reduce its complexity, ChebyNet \cite{defferrard2016convolutional} uses the Chebyshev polynomial to approximate the convolution kernel of the Spectral Graph CNN. After that, Graph Convolution Network (GCN) \cite{kipf2016semi} proposes a semi-supervised method to learn representations of nodes and achieve progress in classification tasks. {In addition}, GraphSAGE \cite{hamilton2017inductive} and Graph Attention Network (GAT) \cite{velivckovic2017graph} use spatial methods to aggregate neighbors' representations of nodes to learn {the embedding of nodes} in different ways. Apart from these graph neural networks mentioned above, some general frameworks of GNNs are proposed to integrate the Graph Neural Networks into one framework. The message-passing neural networks (MPNN) summarize the framework of GNNs into two phases: a message-passing {phase} and a readout phase. Based on this, some platforms {\cite{wu2020comprehensive}} are developed to realize the GNNs. Most of these methods are not designed for anomaly detection, but simply treat it as a classification task without considering the small percentage of anomalous nodes and problems of lacking labels.

In view of the spectrum, the most widely used graph convolution model, GCN, is a low-pass filter to process {signals of nodes} in networks. After that, many models such as SGC \cite{wu2019simplifying}, GWNN \cite{xu2019graph} inherit this idea and propose some graph low-pass filters as the framework of GNNs. {However}, these methods cannot work well because they filter most high-frequency anomalous signals. Therefore, we cannot adopt the GNN-based model in this task directly and we need to design a new framework to adapt.

\section{Preliminaries\label{Preliminaries}}

In this section, we provide definitions of the multi-view attributed networks and formalize the problem of anomaly detection on multi-view attributed networks. The main symbols of this paper are shown in Table \ref{tab:symbels}.

\begin{definition}
    \label{def:one}
    \textbf{Multi-view Attributed Networks} Multi-view attributed networks, denoted as $\mathcal{G} = \{\mathcal{G}^{(\varphi_1)}, \mathcal{G}^{(\varphi_2)},...,\mathcal{G}^{(\varphi_K)}\}$, consist of $n+m$ nodes in $K$ views, where a sub-network $\mathcal{G}^{(\varphi_k)}(k=1, 2,...,K)$ based on a specific view represents a kind of interaction between users and items.
\end{definition}

\begin{table}[]
    \centering
    \caption{Description of key symbols}
    \begin{tabular}{cl}
    \bottomrule[1pt]
    \textbf{Symbols} & \textbf{Definitions} \\
    \hline
    $\mathcal{G}$          &    Multi-view attributed networks  \\
    ${\varphi_k}$          &    The $k$-th view {on} multi-view attributed networks\\
    $\mathcal{G}^{(\varphi_k)}$          &    The sub-network in view $\varphi_k$  \\
    $\mathbf{A}^{(\varphi_k)}\in \mathbb{R}^{n\times n}$          &    The adjacency matrix in view $\varphi_k$  \\
    $\mathbf{X}\in \mathbb{R}^{n\times d}$           &       Attribute matrix of nodes                \\
    $\mathbf{H}^{(l, \varphi_{k})}\in \mathbb{R}^{n \times F_l}$                 &      Attribute matrix of view $\varphi_k$ in layer $l$\\
    $\mathbf{Z}^{(\varphi_k)} \in \mathbb{R}^{n\times F_L}$                 &      Embedding matrix of view $\varphi_k$ \\
    $\mathbf{z}_{i}^{(\varphi_{k})}\in \mathbb{R}^{F_L}$                 &      Embedding vector of node $i$ in view $\varphi_k$ \\
    $\mathcal{L}_{s}^{(\varphi_{k})}$                   & The structure reconstruction loss in view $\varphi_k$ \\
    $\mathcal{L}_{a}$                   & The attribute reconstruction loss \\
    \bottomrule[1pt]
    \end{tabular}
    \label{tab:symbels}
\end{table}
\begin{example}
    In Figure~\ref{fig:toy_example}(a), we defined three interaction modes in shopping networks and they can be treated as multi-view attributed networks $\mathcal{G}$. Customers and goods represent users and items respectively in networks. According to three different interaction modes, they can be treated as multi-view attributed networks with $K=3$, which represent purchasing, making comments, and adding goods into the cart respectively. Figure~\ref{fig:toy_example}(b) shows an example of an interaction action, which represents the co-comment relationship and can be treated as a sub-network $\mathcal{G}^{(\varphi_k)}$ in the view of the co-comment relationship.
\end{example}

The sub-network $\mathcal{G}^{(\varphi_k)}$ of view $\varphi_{k}$ can be defined as:

\begin{definition}
    \label{def:two}
    \textbf{Sub-network $\mathcal{G}^{(\varphi_k)}$ Based on a Specific View.}
    In view $\varphi_{k}$, the sub-network $\mathcal{G}^{(\varphi_k)}=(\mathcal{V}_u, \mathcal{V}_i, \mathcal{E}_{k}, \mathcal{X})$ based on a specific view $\varphi_k$ consists of a vertex set $\mathcal{V}_u$ of users, a vertex set $\mathcal{V}_i$ of items, a relationship set $\mathcal{E}_{k}$ and an attribute set $\mathcal{X}$. The vertex set $\mathcal{V}_u=\{v_1, v_2, ..., v_n\}$ of users consists of $n$ user nodes and the vertex set $\mathcal{V}_i=\{v_{n+1}, v_{n+2}, ..., v_{n+m}\}$ of items consists of $m$ item nodes. Meanwhile, the attribute set $\mathcal{X} = \{\mathbf{x}_1, \mathbf{x}_2, ...,\mathbf{x}_n\}$ represents attributes of user nodes and can be represented as matrix $\mathbf{X}=(\mathbf{x}_1^T, \mathbf{x}_2^T, ...,\mathbf{x}_n^T)^T$, where $\mathbf{x}_i^T \in \mathbf{R}^d$ denotes the attribute vector of each node. For each relationship set $\mathcal{E}_{k}$, a link between two user nodes exists if they have the same interaction relation with the same item in view $\varphi_k$. Therefore, the relationship in view $\varphi_k$ among user nodes can be represented by adjacency matrix $\mathbf{A}^{(\varphi_k)}$, where $\mathbf{A}^{(\varphi_k)}_{ij}=1$ means user $v_i$ and $v_j$ are linked with the same item in view $\varphi_k$, otherwise $\mathbf{A}^{(\varphi_k)}_{ij}=0$.
\end{definition}

\begin{example}
    As shown in Figure~\ref{fig:toy_example}(b), in shopping networks,  we extract relationships of co-comment actions (two users comment on the same item) and get the sub-network in the co-comment view. In Figure~\ref{fig:toy_example}(c), there exist links between the pairs of node 1 and node 2, node 1 and node 3, node 2 and node 3 because all these user pairs {make comments} on the same item.
\end{example}

In Definition \ref{def:two}, we can get the relationship adjacency matrices $\{\mathbf{A}^{(\varphi_k)}\}_{k=1}^{K}$ among users in $K$ views and the attribute matrix $\mathbf{X}$. In anomaly detection tasks, we aim to find out the anomalous users (nodes) in networks. Therefore, we use $\mathcal{V}$ to replace the notation user set $\mathcal{V}_u$. We can also use the adjacency matrix $\mathbf{A}^{(\varphi_k)}$ {among} users in $K$ views to represent the relationships {among} users in subsequent sections.

\begin{problem}

Given multi-view attributed networks $\mathcal{G}$, with attribute matrix $\mathbf{X}$ and adjacency matrices $\{\mathbf{A}^{(\varphi_k)}\}_{k=1}^{K}$ from different views of $n$ nodes, the task is to get anomaly scores of these nodes and rank all these nodes according to the degree of abnormality. The nodes differing from most nodes can be ranked higher.
\label{problem}

\end{problem}

\section{{The} Proposed Method\label{Method}}

\subsection{Motivation and Framework Overview \label{sec:motication}}

\begin{figure}
    \centering
    \subfigure[Sketch of spectrum distribution before and after filtering of normal instances]{
        \includegraphics[width=1.0\textwidth]{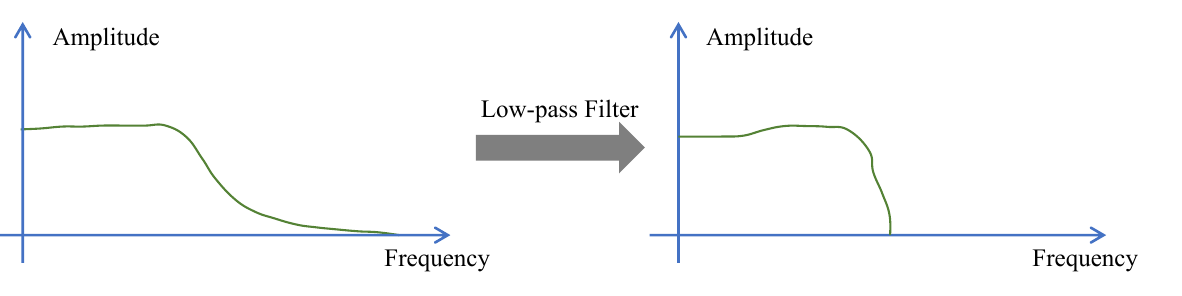}
        \label{fig:draft1}
    }
    \subfigure[Sketch of spectrum distribution before and after filtering of anomalous instances]{
        \includegraphics[width=1.0\textwidth]{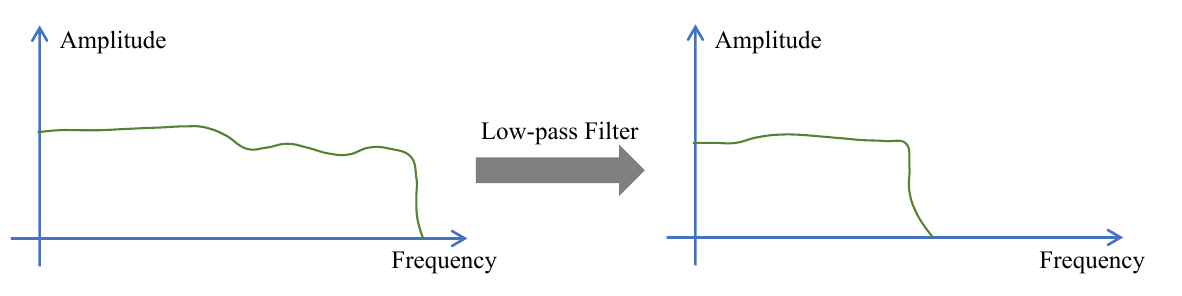}
        \label{fig:draft2}
    }
    \caption{The sketch of the motivation. (a): Normal instances include {few} high-frequency (anomalous) signals. When signals are low-pass filtered, there is a tiny difference in the spectrum distribution before and after {filtering of normal instances}. (b): Anomalous instances include more high-frequency (anomalous) signals than normal ones. When signals are low-pass filtered, there is a larger difference in the spectrum distribution before and after {filtering of anomalous instances}.}
    \label{fig:draft}
\end{figure}

\begin{figure*}
    \centering
    \begin{overpic}[width=0.9\textwidth]{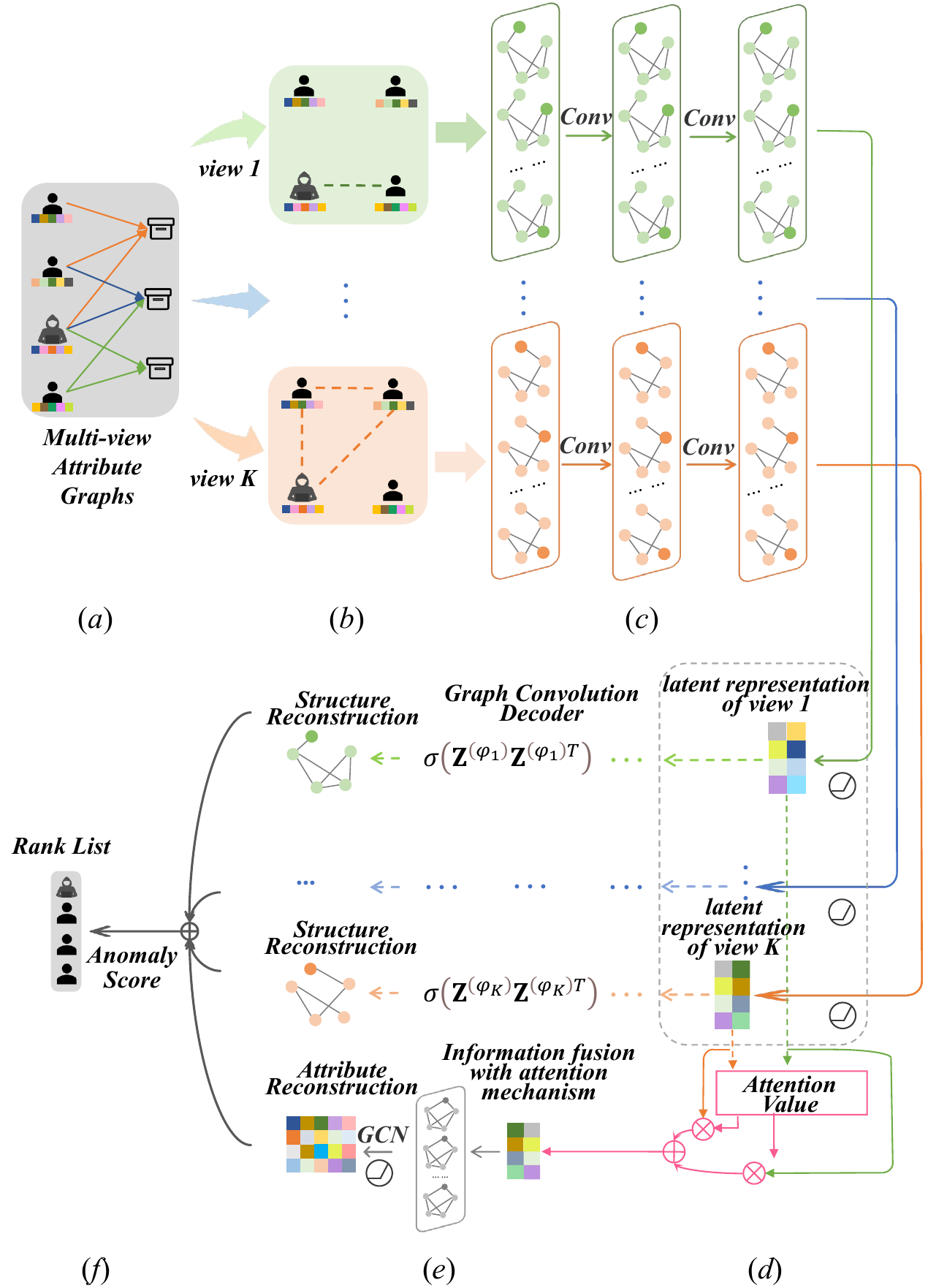}
    \end{overpic}
    \caption{
        The framework of our AnomMAN model.
        (a) Raw input consists of multi-view attributed networks.
        (b) Extract sub-networks based on different views.
        (c) Multi-view attributed {network} encoder layers embed structures and attributes into the same subspace.
        (d) Fuse latent representations of different views together with attention mechanism.
        (e) Reconstruct structures and attributes of nodes.
        (f) Get the anomaly scores according to reconstruction loss.
    }
    \label{tab:modelframework}
\end{figure*}

{First, we introduce our motivation.} To take the attributes and network structure into consideration, we introduce a graph convolution-based framework to detect anomalies. However, the graph convolution operation {serves} as a low-pass filter, which can filter most of the high-frequency (anomalous) signals. {This phenomenon makes it difficult to detect anomalies on multi-view distributed networks.} To overcome this problem, as shown in Figure \ref{fig:draft}, a graph auto-encoder-based method is used. Anomalous instances include more high-frequency signals than normal ones. When signals are low-pass filtered, there is a larger difference in the spectrum distribution before and after filtering. Moreover, this motivation can be more clearly represented by a stronger low-pass filter. Finally, those whose reconstruction errors are larger will be detected as anomalous ones.

In this paper, we propose a model named AnomMAN to detect \textbf{Anoma}ly on \textbf{M}ulti-view \textbf{A}ttributed \textbf{N}etworks. The framework of AnomMAN is shown in Figure~\ref{tab:modelframework}. First, it decomposes multi-view attributed networks $\mathcal{G}$ into $K$ views. Second, it encodes the sub-networks $\mathcal{G}^{(\varphi_k)}$ from each view into the same subspace by multi-view attributed {network} encoder and fuses latent representations from different views with attention mechanism \cite{vaswani2017attention}. Third, we reconstruct both structures and attributes of multi-view attributed networks $\mathcal{G}$. Finally, we get the anomaly score of nodes by both structure and attribute reconstruction errors.

\subsection{Multi-view Attributed {Network} Encoder \label{MANE}}

As shown in Figure \ref{tab:modelframework}, multi-view attributed networks can be divided into $K$ sub-networks by different interactive actions according to Definition \ref{def:one}. In sub-network $\mathcal{G}^{(\varphi_k)}$, a traditional deep auto-encoder is not suitable for processing data with a topological structure. In sub-network $\mathcal{G}^{(\varphi_k)}$, graph convolution \cite{kipf2016semi} layers are used as the encoder.

The inputs of each layer of view $\varphi_{k}$ in multi-view attributed {network} encoder are the adjacency matrix $\mathbf{A}^{(\varphi_k)}$ in view $\varphi_{k}$ and the attribute matrix of nodes $\mathbf{H}^{(l, \varphi_{k})} \in \mathbb{R}^{n \times F_l}$ in each layer, where $n$ and
$F_l$ represent the number of vertexes and the dimension of {attributes of nodes} in layer $l$ respectively. Mathematically, the multi-view attributed {network} encoder is represented as:

\begin{equation}
    \mathbf{H}^{(l+1, \varphi_{k})} = g(\widetilde{\mathbf{A}}^{(\varphi_{k})}\mathbf{H}^{(l, \varphi_{k})}\mathbf{W}^{(l, \varphi_{k})}),
    \label{eq:one}
\end{equation}
where $\mathbf{W}^{(l, \varphi_{k})} \in \mathbb{R}^{F_{l} \times F_{l+1}} $ is the parameter matrix, $g(\cdot)$ is the non-linear activation function, $\mathbf{H}^{(0, \varphi_{k})} = \mathbf{X}$ is the initial attribute matrix and $\widetilde{\mathbf{A}}^{(\varphi_{k})}$ is the adjacency matrix that captures the structure of sub-network $\mathcal{G}^{(\varphi_k)}$ in view $\varphi_{k}$. Equation~\ref{eq:one} denotes the graph convolution propagation of layer $l$ in the multi-view attributed {network} encoder.

The normalized adjacency matrix $\widetilde{\mathbf{A}}^{(\varphi_{k})}$ in view $\varphi_k$ can be calculated as:
\begin{equation}
    \widetilde{\mathbf{A}}^{(\varphi_{k})} = \widetilde{\mathbf{D}}^{{(\varphi_{k})} {-1/2} }
    {\hat{A}}^{(\varphi_{k})}\widetilde{\mathbf{D}}^{{(\varphi_{k})} {-1/2} },
    \label{eq:two}
\end{equation}
where $\hat{\mathbf{A}}^{(\varphi_{k})}=\mathbf{A}^{(\varphi_{k})} + \mathbf{I}$, $\mathbf{I}$ is the identity matrix to add self-loop on each node and $\widetilde{\mathbf{D}}^{(\varphi_{k})}$ is the diagonal degree matrix that satisfies $\widetilde{\mathbf{D}}^{(\varphi_{k})}_{i,i}=\sum_{j}\hat{\mathbf{A}}_{i,j}^{(\varphi_{k})}$.

According to Equations \ref{def:one} and \ref{def:two}, attributes of all nodes in $K$ views are projected to the same algebraic space. Then, we can get the latent representation set of attribute matrices in all views: $\mathbf{Z} = \{\mathbf{Z}^{(\varphi_1)}, \mathbf{Z}^{(\varphi_2)},...,\mathbf{Z}^{(\varphi_K)}\}$, where each matrix $\mathbf{Z}^{(\varphi_k)} \in \mathbb{R}^{n\times F_L}$ and $L$ is the number of the encoding layers. We can denote the latent representation of node $i$ in view $\varphi_k$ as ${\mathbf{z}}^{(\varphi_{k})}_i \in \mathbb{R}^{F_L}(i = 1, 2,..., n)$.

As mentioned above, we use the graph convolution layers as the encoder of multi-view attributed networks in AnomMAN model. However, the graph convolution operation $\widetilde{\mathbf{A}}^{(\varphi_{k})}\mathbf{H}^{(l, \varphi_{k})}$ shown in Equation \ref{eq:one} has been proved to be a low-pass filter in \cite{kipf2016semi, wu2019simplifying}. As described in Section \ref{sec:motication}, in the anomaly detection task, this low-pass filter will filter the high-frequency (anomalous) signals, which will make the application of GCN in graph anomaly detection tasks challenging.
Therefore, we propose a graph auto-encoder-based framework to overcome it. Since most of the anomalous (high-frequency) signals are filtered, the error of anomalous ones between the input and output of graph auto-encoder is larger than non-anomalous ones. Based on this idea, we apply a more powerful low-pass filter encoder to replace the graph convolution encoder formulated in Equation~\ref{eq:one}.
It is hypothesized that the non-linear activation function between two GCN layers is not critical to the graph convolution encode part. Therefore, non-linear activation functions $g(\cdot)$ {among} layers are removed, only remaining the final activation function of the multi-view attributed {network} encoder. The resulting encoder part can be formulated as:
\begin{equation}
    \mathbf{Z}^{(\varphi_{k})} = g( \underbrace{\widetilde{\mathbf{A}}^{(\varphi_{k})}...\widetilde{\mathbf{A}}^{(\varphi_{k})}}_{multiply\ L \  \widetilde{\mathbf{A}}^{(\varphi_{k})}}\mathbf{H}^{(l, \varphi_{k})}\underbrace{\mathbf{W}^{(1, \varphi_{k})}...\mathbf{W}^{(L, \varphi_{k})}}_{multiply \  L \ \mathbf{W}^{(l, \varphi_{k})}}).
    \label{eq:simple}
\end{equation}
Equation~\ref{eq:simple} presents a simplified form of $L$-layer graph convolution layers. The linear transformation matrix $\mathbf{W}^{(l, \varphi_{k})}$ $L$ multiplied by {$L$} times can be replaced by a one-time linear transformation, which improves computational efficiency.

To simplify notations in view $\varphi_k$, we replace the repeated multiplication of the normalized adjacency matrix $\widetilde{\mathbf{A}}^{(\varphi_{k})}$ with a single matrix $(\widetilde{\mathbf{A}}^{(\varphi_{k})})^{L}$. Note that all parameters $\{\mathbf{W}^{(l, \varphi_{k})}\}_{l=1}^{L}$ in Equation~\ref{eq:simple} are learnable by backward propagation algorithm \cite{rumelhart1986learning}. Therefore, it can be reparameterized as:

\begin{equation}
    \mathbf{W}^{(\varphi_{k})} = \mathbf{W}^{(1, \varphi_{k})}\mathbf{W}^{(2, \varphi_{k})}...\mathbf{W}^{(L, \varphi_{k})}.
\end{equation}

\begin{figure}
    \centering
    \subfigure[Multi-layer graph convolution encoder]{
        \includegraphics[width=0.78\textwidth]{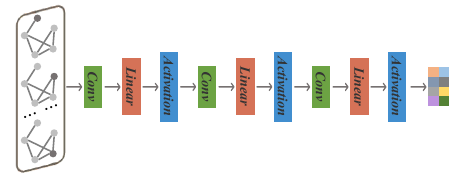}
        \label{fig:sgca}
    }
    \subfigure[Simplified graph convolution encoder]{
        \includegraphics[width=0.78\textwidth]{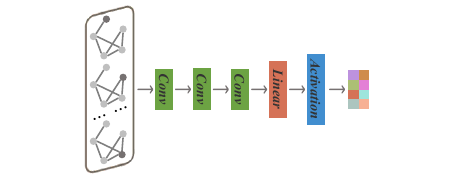}
        \label{fig:sgcb}
    }
    \caption{Differences between multi-Layers graph convolution {encoder} and simplified graph convolution encoder.}
    \label{fig:compare}
\end{figure}

{So far, the simplified multi-view attribute network encoder obtained in view $\varphi_{k}$ can be expressed as:}

\begin{equation}
    \mathbf{Z}^{(\varphi_{k})} = g((\widetilde{\mathbf{A}}^{(\varphi_{k})})^{L}\mathbf{X}\mathbf{W}^{(\varphi_{k})}),
    \label{eq:SGC}
\end{equation}
which is a simpler form of GCN. The contrast between {the} multi-layer graph convolution {encoder} (in Equation~\ref{eq:one}) and {the} simplified graph convolution encoder (in Equation~\ref{eq:SGC}) is shown in Figure \ref{fig:sgca} and Figure \ref{fig:sgcb} respectively.

Now we can analyze why the simplified $L$-order graph convolution Layer shown in Figure \ref{fig:compare} is a more powerful low-pass filter than traditional {$L$-layer} GCN. The decomposition of the Laplacian matrix $L^{(\varphi_k)}$ in a specific view $\varphi_k$ can be formulated as: 
\begin{equation}
L^{(\varphi_k)} = U^{(\varphi_k)}\Lambda^{(\varphi_k)}U^{(\varphi_k)T}, 
\end{equation}
where $U^{(\varphi_k)}$ is the orthogonal matrix composed of eigenvectors of matrix $L^{(\varphi_k)}$ in view and $\Lambda^{(\varphi_k)}$ is the diagonal matrix representing eigenvalues of matrix $L^{(\varphi_k)}$. In the decomposition of the Laplacian matrix, the frequency response function matrix is $I-\Lambda^{(\varphi_k)}$, {which} can be written as $1-\lambda_i^{\varphi_k}$ in an element-wise form.

In simplified graph convolution layer, the convolution kernel $(\widetilde{\mathbf{A}}^{(\varphi_{k})})^{L}$ can be decomposed as:
\begin{equation}
    \begin{aligned}
    (\widetilde{\mathbf{A}}^{(\varphi_{k})})^{L} &= (U^{(\varphi_k)}(I-\Lambda^{(\varphi_k)})U^{(\varphi_k)T})^L
    =U^{(\varphi_k)}(I-\Lambda^{(\varphi_k)})^LU^{(\varphi_k)T},
    \end{aligned}
    \label{eq:simpleconv}
\end{equation}
where the matrix $U^{(\varphi_k)}$ is the matrix of Fourier bases and $(I-\Lambda^{(\varphi_k)})^L$ is the frequency response function of the simplified graph convolution kernel.

In order to find the properties of the graph filter, we need to analyze the characteristic of the frequency response function. In \cite{wu2019simplifying}, it verified that the maximum value of the eigenvalues (known as the frequency in graph convolution operation) of the Laplacian matrix is smaller than 2 and {and generally it is not more than 1.5}, which has also been verified in Section \ref{experiment}. Therefore, we will not consider eigenvalues close to 2. The {element-wise} frequency response function of the simplified encoder with $L$ times' convolution operation is shown in Figure \ref{fig:decompose} (a), and the {element-wise} frequency response function can be formulated as:
\begin{equation}
    f^{L}(\lambda_i^{(\varphi_k)}) = (1-\lambda_i^{(\varphi_k)})^{L}.
\end{equation} In Figure \ref{fig:decompose} (a), the amplitude of the frequency response function is near 0 when the frequency is in the interval (0.5, 1.5), and it has a larger amplitude when the frequency is lower than 1.5. This phenomenon becomes more obvious when the number $L$ of encoding layers is larger. And it reveals that when $L$ is larger, it is a more powerful low-pass filter than traditional multi-layer convolution layers because most low-frequency signals are retained, and most high-frequency signals are filtered.

Therefore, in subsequent sections, the multi-layers introduced in Equation \ref{eq:one} are replaced by the more powerful and simpler low-pass filter in Equation \ref{eq:SGC} as the multi-view attributed {network} encoder.

\subsection{Information Fusion on Multi-view Attributed Networks \label{sec:attentionfusion}}
The latent representation set $\mathbf{Z}$ captures different {behaviors} of users in different views according to the multi-view attributed {network} encoder. Generally, each node {on} multi-view attributed networks contains various kinds of semantic information, and view-specific node representations can only reflect the information of nodes from a specific view. To learn a more comprehensive representation, we fuse representations from all views together. Therefore, we consider relationships among all these views and define the importance of them to get a uniform representation of networks.

In this paper, we adopt the attention mechanism to fuse the latent representations from all views together. According to each latent representation ${\mathbf{z}}^{(\varphi_{k})}_i$ of node $v_i$ in view ${\varphi_{k}}$, the representation of node $v_i$ after the information fusion operation can be formulated as:
\begin{equation}
    \tilde{\mathbf{z}}_i = \sum_{k=1}^{K} \alpha^{(\varphi_k)} {\mathbf{z}}^{(\varphi_{k})}_i,
    \label{eq:three}
\end{equation}
where $\alpha^{(\varphi_k)}$ is the attention weight in view $\varphi_k$. Equation~\ref{eq:three} shows the linear aggregation of different latents from different views. Taking $K$ groups of view-specific node representations learned from multi-view attributed {network} encoder as input, the learned weights of each view ($\alpha^{(\varphi_1)},  \alpha^{(\varphi_2)}, ..., \alpha^{(\varphi_K)}$) can be shown as follows:
\begin{equation}
    (\alpha^{(\varphi_1)}, \alpha^{(\varphi_2)}, ..., \alpha^{(\varphi_K)}) =
    att(\mathbf{Z}^{(\varphi_1)}, \mathbf{Z}^{(\varphi_2)}, ..., \mathbf{Z}^{(\varphi_K)}){,}
    \label{eq:four}
\end{equation}
{where} $att(\cdot)$ operation denotes a deep neural network which performs the attention. It shows that the attention mechanism can capture different kinds of information behind the multi-view attributed networks.

To learn the importance of latent representations in each view, we first adopt a one-layer MLP as a non-linear transformation to transform these representations. {Then}, we measure the importance of latent representations in all $K$ views as the similarity among transformed representations with an attention vector $\mathbf{q}$. {In addition}, we average the importance of the view-specific node {embedding}, which can be explained as the importance of each specific view. Given the latent representations $\mathbf{z}^{(\varphi_1)}_i, \mathbf{z}^{(\varphi_2)}_i, ... ,\mathbf{z}^{(\varphi_K)}_i$ of node $v_i$ from views $\varphi_1, \varphi_2,...,\varphi_K$, the importance of latent representations based on view $\varphi_k$ can be formulated as:
\begin{equation}
    \begin{aligned}
        e^{(\varphi_k)}
            = \frac{1}{|\mathcal{V}|}\sum_{v_i \in \mathcal{V}}\mathbf{q}^{T} \cdot tanh(\mathbf{W}\cdot \mathbf{z}^{(\varphi_k)}_{i}+\mathbf{b}),
    \end{aligned}
    \label{eq:five}
\end{equation}
where $tanh(\cdot)$ is the non-linear activation function, $\mathbf{W}$ is the weight matrix, $\mathbf{b}$ is the bias vector, $\mathbf{q}$ is the attention vector, and $|\mathcal{V}|$ is the number of nodes in the network. Note that all above parameters are shared by all views {on} multi-view attributed networks. In Equation~\ref{eq:five}, $e^{(\varphi_k)}$ implies the contribution of $\mathbf{z}^{(\varphi_k)}_{i}$ in view $\varphi_k$.

According to the importance of the latent representation $e^{(\varphi_j)}_i$, we normalize them with softmax function and the attention weight $\alpha^{(\varphi_j)}_i$ can be formulated as:
\begin{equation}
    \begin{aligned}
        \alpha^{(\varphi_k)} =&
        softmax(e^{(\varphi_k)}),  
    \end{aligned}
    \label{eq:six}
\end{equation}
which can be interpreted as the contribution of view $\varphi_k$ for a specific task, and the $softmax(\cdot)$ operation achieves $\sum_{k=1}^{K} \alpha^{(\varphi_k)}=1$. In Equation~\ref{eq:six}, it can be found that the attention value $\alpha^{(\varphi_j)}_u$ defines different importance of different views, which is asymmetric. According to the latent representation after information fusion, we obtain the new latent representation matrix:
\begin{equation}
    \widetilde{\mathbf{Z}} = (\tilde{\mathbf{z}}_1^T, \tilde{\mathbf{z}}_2^T, ... \tilde{\mathbf{z}}_n^T )^T,
\end{equation} where $\tilde{ \mathbf{z}}_i \in \mathbb{R}^{F_N}$.

To represent in the form of matrix calculation, Equation~\ref{eq:three} can be reformulated as:
\begin{equation}
    \widetilde{\mathbf{Z}} = \sum_{k=1}^{K} \alpha^{(\varphi_k)} {\mathbf{Z}}^{(\varphi_{k})},
    \label{eq:seven}
\end{equation}
which provides a matrix calculation form of Equation~\ref{eq:three}.

\subsection{Attribute and Structure Reconstruction Decoder on Multi-view Attributed Networks\label{sec:decoder}}
After encoding latent representations in different views via multi-view attributed {network} encoder and fusing representations of them together, we need to reconstruct node attributes and network structures in each view.

\subsubsection{Network Structure Reconstruction Decoder}

{On} multi-view attributed networks $\mathcal{G}$, all sub-networks in all views are reconstructed. In other words, for a specific sub-network $\mathcal{G}^{(\varphi_k)}$, structure reconstruction means reconstructing the adjacency matrix in each view. To get the linked probability of two nodes, we first get the inner-product value between two node representations and use the {sigmoid} function to project the value into the interval $(0, 1)$. The probability that node $v_i$ and node $v_j$ are linked can be calculated as:
\begin{equation}
    p(\hat{\mathbf{A}}_{i,j}^{(\varphi_k)}= 1|\mathbf{z}_{i}^{(\varphi_k)}, \mathbf{z}_{j}^{(\varphi_k)})
    =\sigma(\mathbf{z}_{i}^{(\varphi_k)T}\cdot \mathbf{z}_{j}^{(\varphi_k)}),
    \label{eq:eight}
\end{equation}
where $\sigma(\cdot): \mathbb{R} \to (0, 1)$ is the sigmoid function.

Accordingly, the reconstructed adjacency matrix $\hat{\mathbf{A}}^{(\varphi_k)}$ can
be calculated as:
\begin{equation}
    \hat{\mathbf{A}}^{(\varphi_k)} = \sigma(\mathbf{Z}^{(\varphi_k)}\cdot\mathbf{Z}^{(\varphi_k)T}),
    \label{eq:nine}
\end{equation}
which provides a matrix calculation form for network reconstruction calculation.

According to Equation \ref{eq:nine}, we can get the structure reconstruction error: 
\begin{equation}
\mathbf{L}_s^{(\varphi_{k})}=\hat{\mathbf{A}}^{(\varphi_{k})} - \mathbf{A}^{(\varphi_{k})}.
\end{equation} 
Specifically, for a specific node representation in view $\varphi_k$, if its relationship structure can be approximated by the network structure reconstruction decoder, the probability of becoming an anomalous instance is very low.

Through the structure reconstruction error $\mathbf{L}_s^{(\varphi_{k})}$ defined above, we can calculate the structure reconstruction loss. Due to the sparsity of the adjacency matrix, we use the L1-norm to measure it:
\begin{equation}
    \mathcal{L}_{s}^{(\varphi_{k})} = ||\hat{\mathbf{A}}^{(\varphi_{k})} - \mathbf{A}^{(\varphi_{k})}||_{1},
    \label{eq:ten}
\end{equation}
where $||\cdot||_1$ operation denotes the L1-norm of a matrix. It returns the network {structure} reconstruction error in the multi-view attribute networks.

\subsubsection{Network Attribute Reconstruction Decoder }

Similarly, we need to reconstruct the {attributes of nodes}  by a network attribute reconstruction decoder with the latent representation matrix $\widetilde{\mathbf{Z}}$ after information fusion:
\begin{equation}
    \hat{\mathbf{X}}
    = g(\widetilde{\mathbf{D}}^{{-1/2} } \widetilde{\mathbf{A}}\widetilde{\mathbf{D}}^{-1/2} \widetilde{\mathbf{Z}} \mathbf{W}),
    \label{eq:eleven}
\end{equation}
where $\widetilde{\mathbf{A}} \in \mathbb{R}^{n\times n}$ and $\widetilde{\mathbf{D}} \in \mathbb{R}^{n\times n}$ are the normalized adjacency matrix and diagonal degree matrix without considering the various kinds of different views.

In the network attribute reconstruction decoder, the error between attributes and reconstruction attributes is minimized. Therefore, the attribute reconstruction error can be defined as:
\begin{equation}
    \mathbf{L}_{a}=\hat{\mathbf{X}} - \mathbf{X}.
\end{equation}Then, the attribute reconstruction loss function can be formulated as:
\begin{equation}
    \mathcal{L}_{a} = ||\hat{\mathbf{X}} - \mathbf{X}||^{2}_{F},
    \label{eq:twelve}
\end{equation}
where the $||\cdot||^{2}_{F}$ operation denotes the Frobenius norm of the reconstruction error matrix $\mathbf{L}_{a}$.

{So far}, we have introduced the way to reconstruct structures and attributes of multi-view attributed networks. And we obtained the loss of them.

Considering structures and attribute errors jointly, the distance between the original data and the reconstructed data is a powerful indicator for detecting anomalies {on} datasets. Besides, as most anomalous (high-frequency) signals are filtered, instances with a larger error between the input and output of the graph auto-encoder-based framework are considered to be more anomalous ones. Therefore, we take both structure reconstruction loss and attribute reconstruction loss into consideration. The loss function of our model can be
formulated as:

\begin{equation}
    \begin{aligned}
        \mathcal{L} =& \epsilon \mathcal{L}_s + (1-\epsilon)\mathcal{L}_a
        =\epsilon \cdot\frac{1}{K}\sum_{k=1}^{K}\mathcal{L}_s^{(\varphi_k)} + (1-\epsilon)\cdot \mathcal{L}_a,
    \end{aligned}
    \label{eq:thirteen}
\end{equation}
where $\epsilon \in (0,1)$ is a hyper-parameter to balance the importance between structure reconstruction loss and attribute reconstruction loss. A larger value of $\epsilon$ indicates the {greater} importance of the structure of networks anomalies and a small one means the {greater} importance of attributes. According to Equation~\ref{eq:thirteen}, the model needs to minimize the loss function to optimize the parameters of the AnomMAN model.

\subsection{Anomaly Detection \label{sec:anomdect}}

According to the definition in Problem~\ref{problem}, we need to find out nodes with top anomaly scores as anomalous entities. For each node, we adopt the reconstruction loss as its anomaly score:
\begin{equation}
    \begin{aligned}
        \mathbf{Anom}_{i}
        =& \epsilon \cdot \frac{1}{K}\sum_{k=1}^{K}||\hat{\mathbf{a}}_i^{(\varphi_{k})} - \mathbf{a}_i^{(\varphi_{k})}||_{1}
        + (1-\epsilon) \cdot ||\hat{\mathbf{x}}_i - \mathbf{x}_i||^{2}_{2}.
    \end{aligned}
    \label{eq:fourteen}
\end{equation}
where $||\cdot||_1$ means the L1-norm of a vector, $||\cdot||_2$ means the Euclidean norm of a vector, vector $\mathbf{a}_i^{(\varphi_{k})}$ and $\hat{\mathbf{a}}_i^{(\varphi_{k})}$ are the $i$-th row of the matrix $ \mathbf{A}^{(\varphi_{k})}$and $\hat{\mathbf{A}}^{(\varphi_{k})}$ of node $v_i$ to calculate the structure error. With the anomaly score defined in Equation~\ref{eq:fourteen}, the anomaly score list of all nodes  can be obtained and  anomalous ones with higher anomaly scores can be found out.

The pseudocode of the training and anomaly detection procedure of AnomMAN is described in detail in Algorithm~\ref{alg:algorithm}. With the Algorithm~\ref{alg:algorithm}, we can get anomaly scores of these nodes. Then, we rank all these nodes according to the degree of abnormality. The nodes differing from most nodes (with higher anomaly scores) can be ranked higher.

\begin{algorithm}
    \caption{The Framework of AnomMAN}
    \KwIn{Multi-view attributed networks $\mathcal{G}= \{\mathcal{G}^{(\varphi_1)}, \mathcal{G}^{(\varphi_2)},$ $...,\mathcal{G}^{(\varphi_K)}\}$ having $K$ views and in each view $\mathcal{G}^{(\varphi_k)}$ $=(\mathcal{V}, \mathcal{E}_{k}, \mathcal{X})$. Maximum iterations: $MaxIter$.}
    \KwOut{Anomaly score of nodes.}
    \For{$iter \in 0, 1,..., MaxIter$}{
        \For{$k \in 1,2,...,K$}{
            Encode $\mathcal{G}^{(\varphi_k)}$ of each view via Equation~\ref{eq:one} and get the latent representation $\mathbf{Z}^{(\varphi_k)}$ of it\;
        }
        Fuse latent representations of different views to $\widetilde{\mathbf{Z}}$ via Equation~\ref{eq:three},\ref{eq:four} and \ref{eq:five}\;
        Reconstruct both attributes and structures of $\mathcal{G}$\;
        \For{$k \in 1,2,...,K$}{
            Calculate the structure reconstruction loss $\mathcal{L}_s^{(\varphi_k)}$ of each view via Equation \ref{eq:ten} respectively\;
        }
        Calculate attribute reconstruction loss $\mathcal{L}_a$\;
        Calculate loss function of AnomMAN: $\mathcal{L} = \epsilon \mathcal{L}_s + (1-\epsilon)\mathcal{L}_a$ $ = \epsilon \cdot\frac{1}{K}\sum_{k=1}^{K}\mathcal{L}_s^{(\varphi_k)} + (1-\epsilon)\cdot \mathcal{L}_a$\;
        Back propagation the loss function $\mathcal{L}$ and update parameters in AnomMAN\;
    }
    Get anomaly scores of all nodes via Equation~\ref{eq:fourteen}\;
    \textbf{return} Anomaly score of all nodes\;
    \label{alg:algorithm}
\end{algorithm}

\subsection{Analysis of the Proposed Model}
\subsubsection{Theory Analysis on the Low-pass Filter Properties}

One of the main claims in the proposal is taking advantage of the low-pass filter properties. Here we present the theoretical analysis of the low-pass filter properties of the AnomMAN framework. AnomMAN is composed of three core components: 1) multi-view attributed {network} encoder, 2) information fusion via attention aggregation, and 3) attribute and structure reconstruction decoder. The low-pass filter properties of the multi-view attributed {network} encoder and attribute and structure reconstruction decoder are both graph convolution layers, whose low-pass filter properties are well discussed in Section~\ref{MANE}. Here we analyze whether the attention aggregation operation maintains the low-pass filter property between the multi-view attributed {network} encoder and attribute and structure reconstruction decoder.
\begin{proposal}
    {Assume that $U$ and $\Lambda$ matrices of decomposition results of $K$ Laplacian matrices are the same respectively}, w.r.t. $U^{(\varphi_k)} = U$ and $\Lambda^{(\varphi_k)} = \Lambda$ for $i = 1, 2, \cdots, K$. Let $\alpha^{(\varphi_k)}$s denote the attention values in each view. The attention aggregation mechanism $\widetilde{\mathbf{Z}} = \sum_{k=1}^{K} \alpha^{(\varphi_k)} {\mathbf{Z}}^{(\varphi_{k})}$ keeps the low-pass filtering property between the multi-view attributed {network} encoder $\mathbf{Z}^{(\varphi_{k})} \approx (\widetilde{\mathbf{A}}^{(\varphi_{k})})^{L}\mathbf{X}'$ and attribute and structure reconstruction decoder, where $\mathbf{X}' = \mathbf{X}\mathbf{W}$.
\end{proposal}
\begin{proof}
    According to Equation \ref{eq:simpleconv}, the multi-view attributed {network} encoder can be decomposed as:
    \begin{equation}
        \begin{aligned}
            \mathbf{Z}^{(\varphi_{k})}  &\approx(\widetilde{\mathbf{A}}^{(\varphi_{k})})^{L}\mathbf{X}' \\  &=[U^{(\varphi_k)}(I-\Lambda^{(\varphi_k)})U^{(\varphi_k)T}]^L\mathbf{X}' \\  &=U^{(\varphi_k)}(I-\Lambda^{(\varphi_k)})^LU^{(\varphi_k)T}\mathbf{X}' \\  
            &=U(I-\Lambda)^LU^\mathrm{T}\mathbf{X}' .
        \end{aligned}
    \end{equation}
    Therefore, the attention mechanism on $\mathbf{Z}^{(\varphi_{k})}$ can be decomposed as:
    \begin{equation}
        \begin{aligned}
            \widetilde{\mathbf{Z}} &= \sum_{k=1}^{K} \alpha^{(\varphi_k)} {\mathbf{Z}}^{(\varphi_{k})}\\ &=\sum_{k=1}^{K} \alpha^{(\varphi_k)} (\widetilde{\mathbf{A}}^{(\varphi_{k})})^{L}\mathbf{X}'\\ &=  \sum_{k=1}^{K} \alpha^{(\varphi_k)} U(I-\Lambda)^LU^\mathrm{T}\mathbf{X}'\\   
            &=  U [\sum_{k=1}^{K} \alpha^{(\varphi_k)} (I-\Lambda)^L ]U^\mathrm{T}\mathbf{X}'\\   
            &=  \mathcal{F}^{-1}\left \{ H(\Lambda)\cdot \mathcal{F}(\mathbf{X}') \right \},\\ 
        \end{aligned}
    \end{equation}
    where $H(\Lambda)=\sum_{k=1}^{K} \alpha^{(\varphi_k)} (I-\Lambda)^L$, $\mathcal{F}(\cdot)$ and $\mathcal{F}^{-1}(\cdot)$ denotes the Fourier transform and inverse transform respectively. Therefore, the frequency response function can be formulated as:
    \begin{equation}
        H(\Lambda)=\sum_{k=1}^{K} \alpha^{(\varphi_k)} (1-\Lambda)^L=\left ( \sum_{k=1}^{K} \alpha^{(\varphi_k)}\right ) (1-\Lambda)^L = (1-\Lambda)^L,
    \end{equation}
    where $\sum_{k=1}^{K} \alpha^{(\varphi_k)}=1$.
    The frequency response function shows that the attention mechanism keeps the low-pass filter property shown in Section~\ref{MANE}.
As the three core components maintain the low-pass filter properties, the low-pass filter properties are satisfied in our AnomMAN framework.
\end{proof}

\subsubsection{Time Complexity Analysis of AnomMAN}
The proposed AnomMAN is highly efficient and can be easily parallelized. The main computation cost is {caused by} the multi-view attributed {network} encoder and attention aggregation. The times of linear transformation operations in the simplified graph convolution encoder reduce to $\mathcal{O}(1)$ comparing the Multi-layer GCNs with $\mathcal{O}(L)$ times. As for the attention aggregation operation, the computation complexity is $\mathcal{O}(|\mathcal{V}|D_1D_2K)$, where the $|\mathcal{V}|$ denotes the number of nodes, $D_1$ denotes the dimension of the attention vector, $D_2$ denotes the dimension of the latent vectors, and $K$ denotes the number of views. The overall complexity is linear to the number of nodes and the number of views. As the multi-view attributed {network} encoder can be parallelized across $K$ views, the proposed AnomMAN model can be easily parallelized.

\section{Experiments\label{experiment}}

In this section, we perform an empirical evaluation on two real-world datasets to prove the verification of our proposed model AnomMAN.

\subsection{Datasets}

We evaluate our model on two real-world datasets. Due to the shortage of ground truth of anomalies, we inject anomalies into these datasets following the perturbation scheme introduced in \cite{ding2019deep} (including both {structure} anomaly and attribute anomaly {on} multi-view attributed networks). Nodes with both attribute anomaly and {structure} anomaly account for 50 \% of the whole dataset.

\textbf{DBLP}\footnote{\url{https://dblp.uni-trier.de/}}:
This is an academic cooperation multi-view attributed networks from the DBLP dataset. There are 3025 nodes {on} multi-view attributed networks, {and} each node represents a specific author. In this dataset, the number of views is set to $K=3$. The 3 views in networks defined in this dataset include co-authorship (two authors cooperated for the same paper), co-conference (two authors published a paper at the same conference), and co-term (two authors published papers in the same term). Authors’ attributes correspond to elements of a bag-of-words represented by keywords.

\textbf{IMDB}\footnote{\url{https://www.imdb.com/}}
This is a movie relationship multi-view attributed networks from the IMDB dataset. There are 4780 nodes {on} multi-view attributed networks, each node represents a specific movie. In this dataset, the number of views is set to $K=3$. The 3 views defined in this dataset include co-actor (two movies are acted by the same actor), co-director (two movies are directed by the same director), and co-year (two movies are released in the same year). Movie attributes are elements of a bag-of-words represented by plots.

The detailed summary of DBLP and IMDB datasets is shown in Table \ref{tab:datasets}.
\begin{table*}[hbp]
    \tabcolsep=5pt
    \centering
    \caption{Summary of Multi-view Attributed Networks Datasets}
    \begin{tabular}{ccccc}
    \bottomrule[1pt]
    \textbf{Datasets} & \textbf{Nodes} & \textbf{Attributes} & \textbf{Anomaly} & \textbf{Edges in each view} \\
    \hline
    DBLP     & 3025  & 334 & 300  & \makecell[c]{co-authorship (11,113),\\ co-conference (5,000,495),\\ co-term (6,776,335)}  \\
    IMDB     & 4780  & 1232 & 300 & \makecell[c]{co-actor (98,010),\\ co-director (21,018),\\ co-year (813,852)}    \\
    \bottomrule[1pt]
    \end{tabular}
    \label{tab:datasets}
\end{table*}
\subsection{{Method Comparison}}
 The proposed method is compared with several baseline methods from 5 perspectives, and the comparison between baselines and our AnomMAN method is shown in Table \ref{tab:comparsion}. LOF \cite{breunig2000lof} is the model to detect anomalies by context and only considers the attributes of nodes. Based on the residual analysis and CUR decomposition, ANOMALOUS \cite{peng2018anomalous} detects the anomalies on the attributed networks. Radar \cite{li2017radar} detects anomalies by analyzing the residuals of {attributes of nodes}. Dominant \cite{ding2019deep} is the algorithm to detect anomalies on attributed networks, which does not take different views of networks into consideration. Dominant and Radar are two state-of-the-art algorithms to detect anomalies on attributed networks. To summarize, we compare our model AnomMAN with LOF, Radar, ANOMALOUS, and Dominant in detecting anomalies on multi-view attributed networks. In anomaly detection tasks, although some models (like SpotLight \cite{Eswaran2018spotlight} and NetWalk \cite{yu2018netwalk}) show good results, they are all designed for temporal anomaly detection tasks. It is inappropriate to take SpotLight and NetWalk models into the comparison of this paper. To verify our motivation, AnomMAN$_{avg}$ and AnomMAN$_{mul}$ are also introduced in the ablation study. These are two variants of AnomMAN. AnomMAN$_{avg}$ replaces the attention module with the average operation to aggregate the latent representations from each view equally. In other words, it treats the same importance of data from different perspectives. Besides, AnomMAN$_{mul}$ uses traditional multi-layer graph convolution networks (in Equation \ref{eq:one}) to replace the more powerful low-pass filter (in Equation \ref{eq:SGC}) as the encoder. 
\begin{table*}[hp]
    \tabcolsep=5pt
    \centering
    \setlength{\tabcolsep}{2.5mm}{} 
    \caption{Comparsion of Baselines and AnomMAN}
    \begin{tabular}{cccccc}
    \bottomrule[1pt]
     & \multicolumn{1}{p{4.19em}}{\tabincell{c}{\textbf{{Attribute}}}} & \multicolumn{1}{p{4.19em}}{\tabincell{c}{\textbf{Network}\\\textbf{Structure}}}  & \multicolumn{1}{p{5.0em}}{\tabincell{c}{\textbf{Non-Linear}\\\textbf{Model}}} &\multicolumn{1}{p{2.19em}}{\tabincell{c}{\textbf{Multi}\\\textbf{-view}}}
     & \multicolumn{1}{p{4.19em}}{\tabincell{c}{\textbf{Low-pass}\\\textbf{Property}}}\\
    \hline
    LOF      & \CheckmarkBold  & \XSolidBrush & \XSolidBrush  &\XSolidBrush  & \XSolidBrush \\
    ANOMALOUS      & \CheckmarkBold  &  \CheckmarkBold  & \XSolidBrush  & \XSolidBrush  & \XSolidBrush \\
    Radar      & \CheckmarkBold   & \CheckmarkBold  & \XSolidBrush  & \XSolidBrush  & \XSolidBrush \\
    Dominant     & \CheckmarkBold   & \CheckmarkBold   & \CheckmarkBold   & \XSolidBrush  & \XSolidBrush \\
    \textbf{AnomMAN}      & \CheckmarkBold   & \CheckmarkBold   & \CheckmarkBold  & \CheckmarkBold   &  \CheckmarkBold  \\
    \bottomrule[1pt]
    \end{tabular}
    \label{tab:comparsion}
\end{table*}
\begin{table*}
    \tabcolsep=2pt
    \centering
    \caption{Results of Anomaly Detection on Multi-view Networks w.r.t. Precision{@$K$}}
    \begin{tabular}{ccccc|ccccc}
        \bottomrule[1pt]
        \textbf{}                     & \multicolumn{3}{c}{\textbf{DBLP}}                                                         & \multicolumn{1}{l}{\textbf{}} & \multicolumn{3}{c}{\textbf{IMDB}}                                 \\
        \hline
        K                             & 50             & 150            & 200                        & 300                        & K                    & 50             & 150            & 200            & 300            \\
        LOF                           & 0.010          & 0.027          & 0.030                      & 0.047                      & LOF                  & 0.007          & 0.007          & 0.010          & 0.027          \\
        ANOMALOUS                     & 0.003          & 0.007          & 0.010                      & 0.027                      & ANOMALOUS            & 0.007          &0.013           & 0.017          & 0.027          \\
        Radar                         & 0.157          & 0.417          & 0.450                      & 0.490                      & Radar                & 0.123          & 0.300          & 0.313          & 0.347          \\
        Dominant                      & 0.133          & 0.350          & 0.403                      & 0.463                      & Dominant             & 0.160          & 0.447          & 0.457          & 0.483          \\
        AnomMAN$_{avg}$               & \textbf{0.167} & 0.410          & 0.420                      & 0.440                      & AnomMAN$_{avg}$      & 0.153          & 0.463          & 0.473          & 0.487          \\
        AnomMAN$_{mul}$               & \textbf{0.167} & 0.420          & 0.473                      & 0.503                      & AnomMAN$_{mul}$      & 0.147          & 0.480          & 0.487          & 0.507          \\
        \textbf{AnomMAN}              & \textbf{0.167} & \textbf{0.490} & \textbf{0.510}             & \textbf{0.523}             & \textbf{AnomMAN}     &\textbf{0.167}  & \textbf{0.493} & \textbf{0.507} & \textbf{0.530} \\
        \bottomrule[1pt]
    \end{tabular}
    \label{tab:precision}
\end{table*}

\subsection{Experimental Setup and Metrics}

In experiments on DBLP and IMDB datasets, we adopt Adam \cite{kingma2014adam} algorithm to minimize the objective function in Equation \ref{eq:thirteen}. The embedding dimension of the latent representations $\textbf{z}_{i}^{(\varphi_k)}$ is set to 30 and the hyper-parameter $L$ is set to 3 in multi-view attributed {network} encoder. We set the learning rate to 0.001, the parameter $\epsilon$ to 0.5 (more ablation studies in Figure \ref{fig:epsilon}), and use ReLU as the activation function in multi-view attributed {network} encoder. For the attribute-based algorithm (LOF), we only train the model with its attributes. Their interactions are treated as the same. We train our models on the machine with one NVIDIA A100-40GB GPU.
\begin{table*}
    \tabcolsep=2pt
    \centering
    \caption{Results of Anomaly Detection on Multi-view Networks w.r.t. Recall{@$K$}}
    \begin{tabular}{ccccc|ccccc}
        \bottomrule[1pt]
        \textbf{}                & \multicolumn{3}{c}{\textbf{DBLP}}                                        & \multicolumn{1}{l}{\textbf{}} & \multicolumn{3}{c}{\textbf{IMDB}}                        \\
        \hline
        K                        & 50             & 150            & 200            & 300                   & K                    & 50             & 150            & 200            & 300            \\
        LOF                      & 0.060          & 0.053          & 0.045          & 0.047                 & LOF                  & 0.040          & 0.013          & 0.015          & 0.027          \\
        ANOMALOUS                & 0.020          & 0.013          & 0.015          & 0.027                 & ANOMALOUS            & 0.040          & 0.027          & 0.025          & 0.027          \\
        Radar                    & 0.940          & 0.833          & 0.675          & 0.490                 & Radar                & 0.740          & 0.600          & 0.470          & 0.347          \\
        Dominant                 & 0.800          & 0.700          & 0.605          & 0.463                 & Dominant             & 0.960          & 0.893          & 0.685          & 0.483          \\
        AnomMAN$_{avg}$          & \textbf{1.000} & 0.820          & 0.630          & 0.440                 & AnomMAN$_{avg}$      & 0.920          & 0.927          & 0.710          & 0.487          \\
        AnomMAN$_{mul}$          & \textbf{1.000} & 0.840          & 0.710          & 0.503                 & AnomMAN$_{mul}$      & 0.880          & 0.960          & 0.730          & 0.507 \\
        \textbf{AnomMAN}         & \textbf{1.000} & \textbf{0.980}  & \textbf{0.765} & \textbf{0.523}        & \textbf{AnomMAN}     & \textbf{1.000}       & \textbf{0.987} & \textbf{0.760} & \textbf{0.530} \\
        \bottomrule[1pt]
    \end{tabular}
    \label{tab:recall}
\end{table*}

For a fair comparison, all parameters of baseline models are well-tuned. For the LOF algorithm, we set the number of neighbors as 20 and the L2 Minkowski distance as the metric to calculate the distance. For ANOMALOUS, the algorithm runs in 20 iterations with parameters $\alpha = 0.015, \beta = 0.01, \gamma = 0.009$, and $\phi = 0.6$. For Radar, we reproduce the algorithm with parameters $\alpha = 0.1, \beta = 0.01$, and $\gamma = 0.1$. For Dominant, the model is optimized with the Adam algorithm (the learning rate is set as 0.005) and trained for 300 epochs.

\begin{figure}
    \centering
    \subfigure[The ROC curve of AnomMAN on DBLP dataset]{
        \includegraphics[scale=0.4]{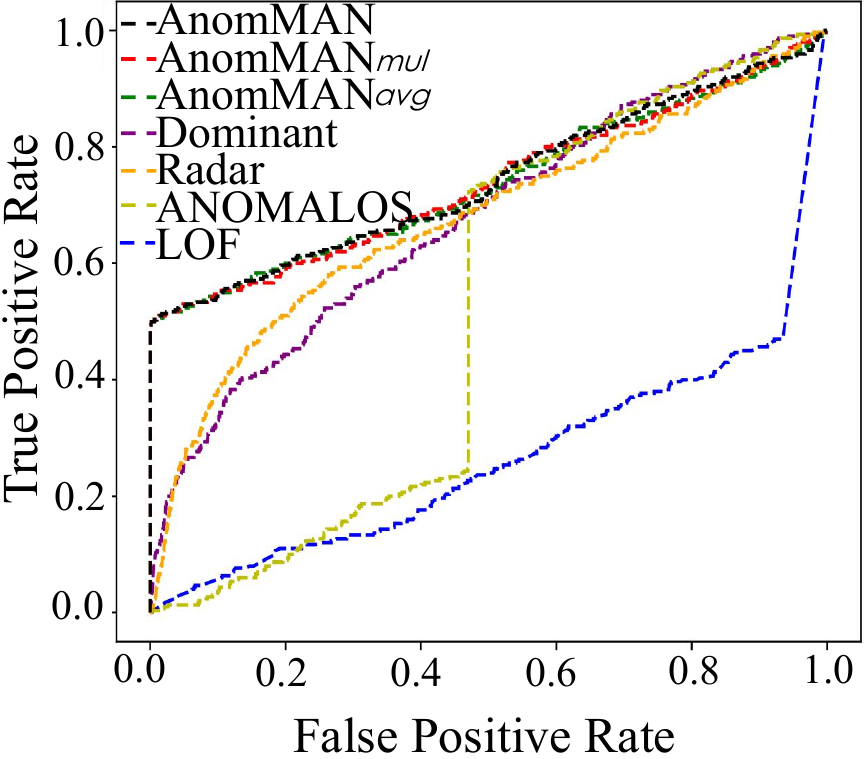}
        \hspace{0mm}
        \label{Fig.sub.roca}
    }
    \subfigure[The ROC curve of AnomMAN on IMDB dataset]{
        \includegraphics[scale=0.4]{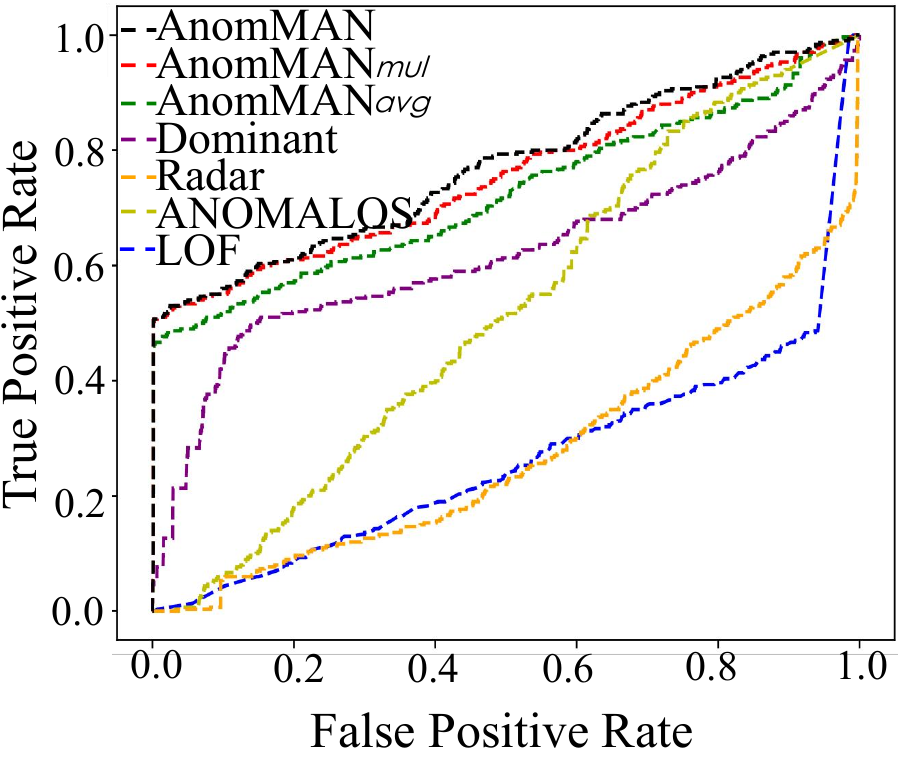}
        \label{Fig.sub.rocb}
    }
    \caption{ROC curve of all methods on two datasets.}
    \label{tab:roc}
\end{figure}

For performance evaluation, we choose metrics widely used in anomaly detection (Precision{@$K$}, Recall{@$K$}, and ROC-AUC) to measure the performances of different algorithms.

The Precision{@$K$} is defined as:

\begin{equation*}
    Precision@K = \frac{\# of\ anomalous\ instances\ in\ the\ Top-K \ list}{\# \ of \ K}.
\end{equation*}

The Recall{@$K$} is defined as:

\begin{equation*}
    Precision@K = \frac{\# of\ anomalous\ instances\ in\ the\ Top-K\ list}{total\ \# \ of\ anomalous\ instances}.
\end{equation*}

{And} the AUC value is the area under the ROC curve.

\subsection{Result Analysis}

In experiments, we compare the proposed model AnomMAN with the aforementioned baselines and the variants of AnomMAN. The Precision{@$K$} and Recall{@$K$} value of all methods are presented in Table \ref{tab:precision} and Table \ref{tab:recall} respectively. {The} AUC value is presented in Table \ref{tab:auc} and the ROC curve in Figure \ref{tab:roc}. The ROC curve of the DBLP dataset and IMDB dataset are shown in Figure \ref{tab:roc}(a) and Figure \ref{tab:roc}(b) respectively. According to the evaluation results, we make some observations as follows:

\begin{table}
    \centering
    \caption{AUC value of Anomaly Detection on Multi-view Attributed Networks}
    \begin{tabular}{c|cc}
    \bottomrule[1pt]
                  & DBLP & IMDB \\
    \hline
    LOF              & 0.265                & 0.264             \\
    ANOMALOUS        & 0.518                & 0.516             \\
    Radar            & 0.678                & 0.275             \\
    Dominant         & 0.677                & 0.621             \\
    AnomMAN$_{avg}$  & 0.737                & 0.722             \\
    AnomMAN$_{mul}$  & 0.738                & 0.756             \\
    \textbf{AnomMAN} & \textbf{0.745}       & \textbf{0.770}    \\
    \bottomrule[1pt]
    \end{tabular}
    \label{tab:auc}
\end{table}

\begin{itemize}
    \item {
        The proposed model AnomMAN outperforms other baseline methods on both DBLP and IMDB datasets. It verifies the effectiveness of the AnomMAN model in detecting anomalies on multi-view attributed networks.
    }
    \item {
        LOF can only detect a limited number of anomalous instances in the network because they cannot overcome the impact of data non-linearity and network sparseness. {And} the residual analysis-based model (ANOMALOUS and Radar) cannot resolve the problem of network sparseness and data non-linearity. All methods above are based on quasi-convex optimization algorithms, and they cannot approximate the distribution of real-world data accurately when facing high-dimension and non-linearity data in a complex, sparse network. Although Dominant shows better performance than other baseline algorithms, it cannot mine the relationships among different views. Moreover, the Recall@50 indicator of AnomMAN even reaches 1.000 on the dataset. It shows that the top 50 anomalous instances detected by AnomMAN are all real anomalous instances.
    }
    \item {
        AnomMAN shows a better performance than baselines according to the results of Precision{@$K$}, Recall{@$K$}, and AUC value. It outperforms other algorithms when getting the anomaly interaction actions from different views and fusing the information. In the ablation study, AnomMAN shows a better performance than AnomMAN$_{avg}$ because it can learn the importance of views to fuse information. As AnomMAN uses a more powerful low-pass filter encoder to filter high-frequency (anomalous) signals, it makes anomalous instances have larger errors between the input and reconstruction information of the framework than in AnomMAN$_{mul}$. Therefore, AnomMAN shows better performance than AnomMAN$_{mul}$.
    }
\end{itemize}

\subsection{Ablation Study on Different Kinds of Norms}
In the objective function and anomaly detection part, we calculate the norm of the matrices in Equations~\ref{eq:twelve} and \ref{eq:fourteen}. Here we show the ablation studies of different kinds of norm calculations on both DBLP and IMDB datasets, and the experimental results are shown in Table~\ref{tab:normablation}. The AUC metric is used to evaluate the performance. In Equations~\ref{eq:twelve} and \ref{eq:fourteen}, we calculate the structural error with L1-norm and the attribute error with L2-norm. We employ different combinations of norms in our experiments, and the last row of Table~\ref{tab:normablation} denotes our proposed AnomMAN Algorithm. Experimental results show that the {structure} errors with L1-norm and the attribute errors with L2-norm perform best among other kinds of combinations. This result is mainly {caused by} the sparsity of networks and the continuity of {attributes of nodes}. This further verifies the reasonability of the model design.
\begin{table}[hbp]
    \centering
    \caption{Ablation Study on Different Kinds of Norms on the AUC Metric}
    \begin{tabular}{cc|cc}
    \bottomrule[1pt]
    Structural Error &  Attribute Error  & DBLP & IMDB \\
    \hline
    L2-norm         &  L1-norm         & 0.725                & 0.751             \\
    L1-norm  &  L1-norm         & 0.742                & 0.760             \\
    L2-norm  &  L2-norm         & 0.739                & 0.762             \\
    L1-norm &  L2-norm         & \textbf{0.745}       & \textbf{0.770}    \\
    \bottomrule[1pt]
    \end{tabular}
    \label{tab:normablation}
\end{table}

\subsection{Parameter Sensitivity Analysis}
\begin{figure}
    \centering
    \includegraphics[width=0.68\textwidth]{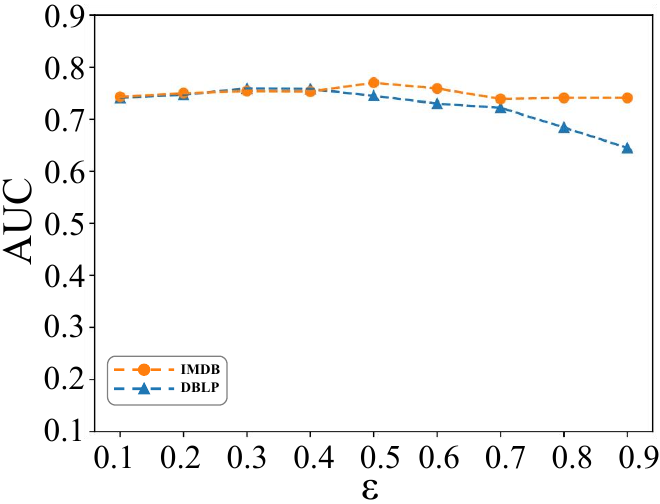}
    \caption{The impact of different $\epsilon$ w.r.t. AUC values.}
    \label{fig:epsilon}
\end{figure}

We also investigate the sensitivity of parameters and report the results of anomaly detection on multi-view attributed networks on both DBLP and IMDB datasets.

\subsubsection{The parameter $\epsilon$ to control the importance between structures and attributes}

The sensitivity of parameter $\epsilon$ on both DBLP and IMDB datasets is first investigated.
The result is shown in Figure \ref{fig:epsilon}.
The parameter $\epsilon$ controls the balance of
the importance between structure reconstruction error
and attribute reconstruction error.
We use the AUC value as a metric to evaluate the sensitivity of parameter $\epsilon$.
The value of $\epsilon$ varies from 0.1 to 0.9
with an interval of 0.1.
The result of the experiments shows that
it is necessary to control the value of $\epsilon$
to balance the structure reconstruction loss
and attribute reconstruction loss
for achieving better performance.
The reasonable value of $\epsilon$ can be chosen
from 0.2 to 0.5 for the DBLP dataset and
from 0.2 to 0.6 for the IMDB dataset.

\begin{figure*}[hp]
    \centering
    \subfigure[The impact of embedding dimension on Recall{@$K$} {on} DBLP dataset]{
        \includegraphics[width=0.45\textwidth]{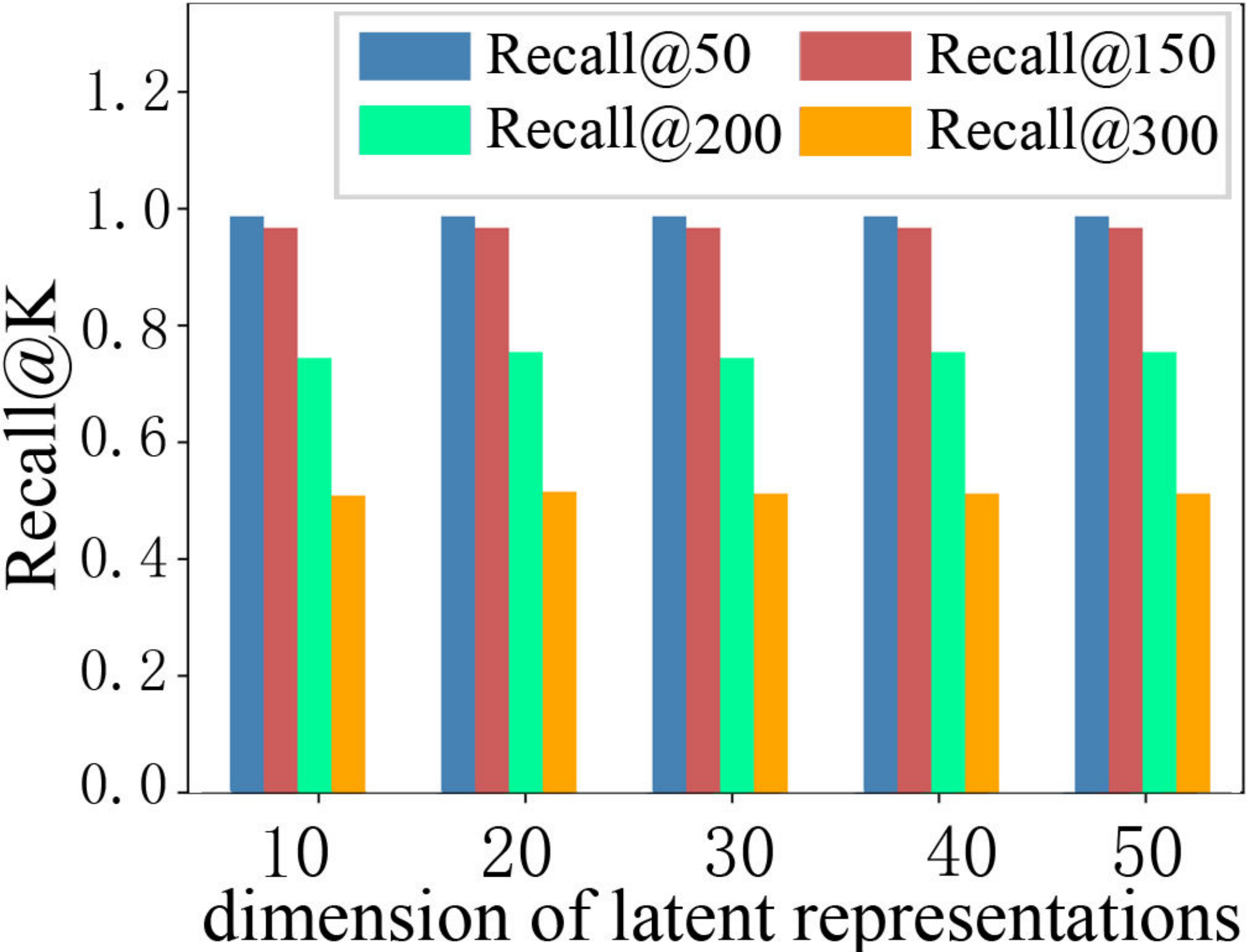}
        \label{Fig.sub.embeddingrecal2}
    }
    \subfigure[The impact of embedding dimension on Recall{@$K$} {on} IMDB dataset]{
        \includegraphics[width=0.45\textwidth]{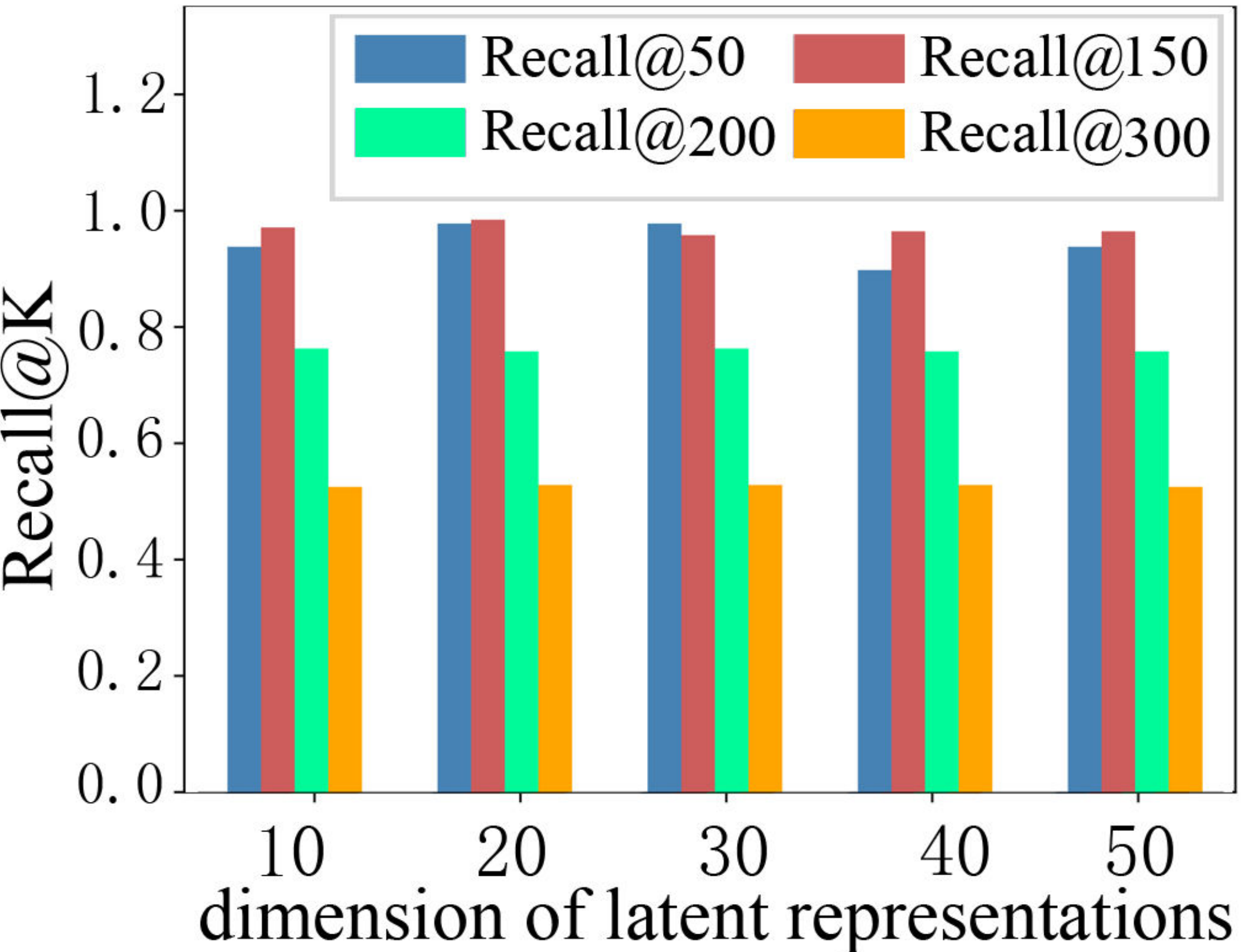}
        \label{Fig.sub.embeddingrecal3}
    }
    \caption{the impact of the {embedding dimension} on both DBLP and IMDB datasets.}
    \label{fig:embeddingrecpre}
\end{figure*}
\subsubsection{The dimension of the latent {representation layers}}

Since the performance of encoder part in AnomMAN framework is related to the dimension of the latent representation matrix $\tilde{\textbf{Z}}$. We chose 5 different choices of dimensions (15, 30, 60, 300, 500) of the latent representation layers on both DBLP and IMDB datasets.

\begin{figure*}[hp]
    \centering
    \subfigure[The impact of embedding dimension on AUC value {on} both DBLP and IMDB datasets]{
        \includegraphics[width=0.45\textwidth]{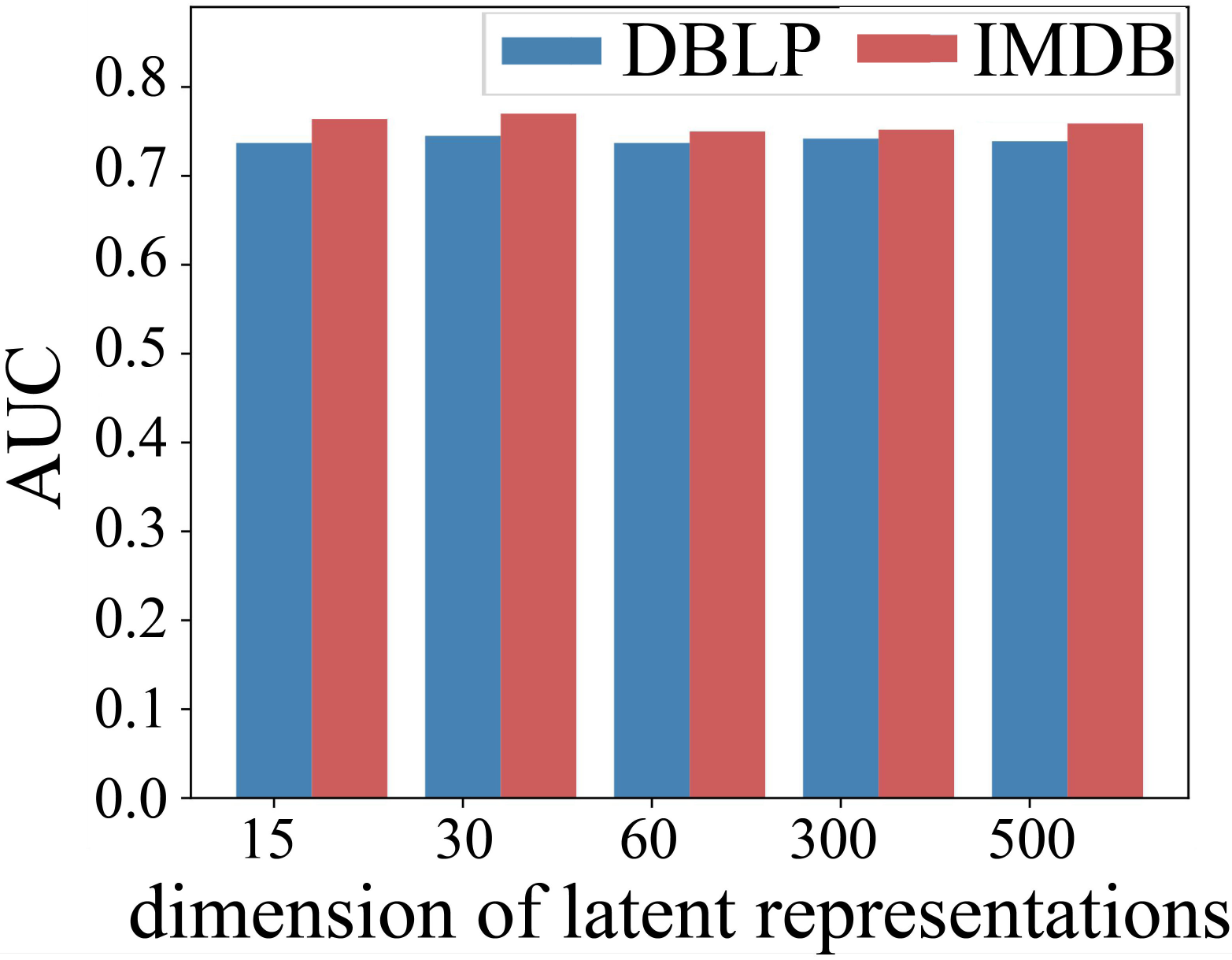}
        \label{Fig.sub.embeddingauc}
    }
    \subfigure[the impact of the dimension of attention vector $\mathbf{q}$ on AUC value {on} both DBLP and IMDB datasets]{
        \includegraphics[width=0.45\textwidth]{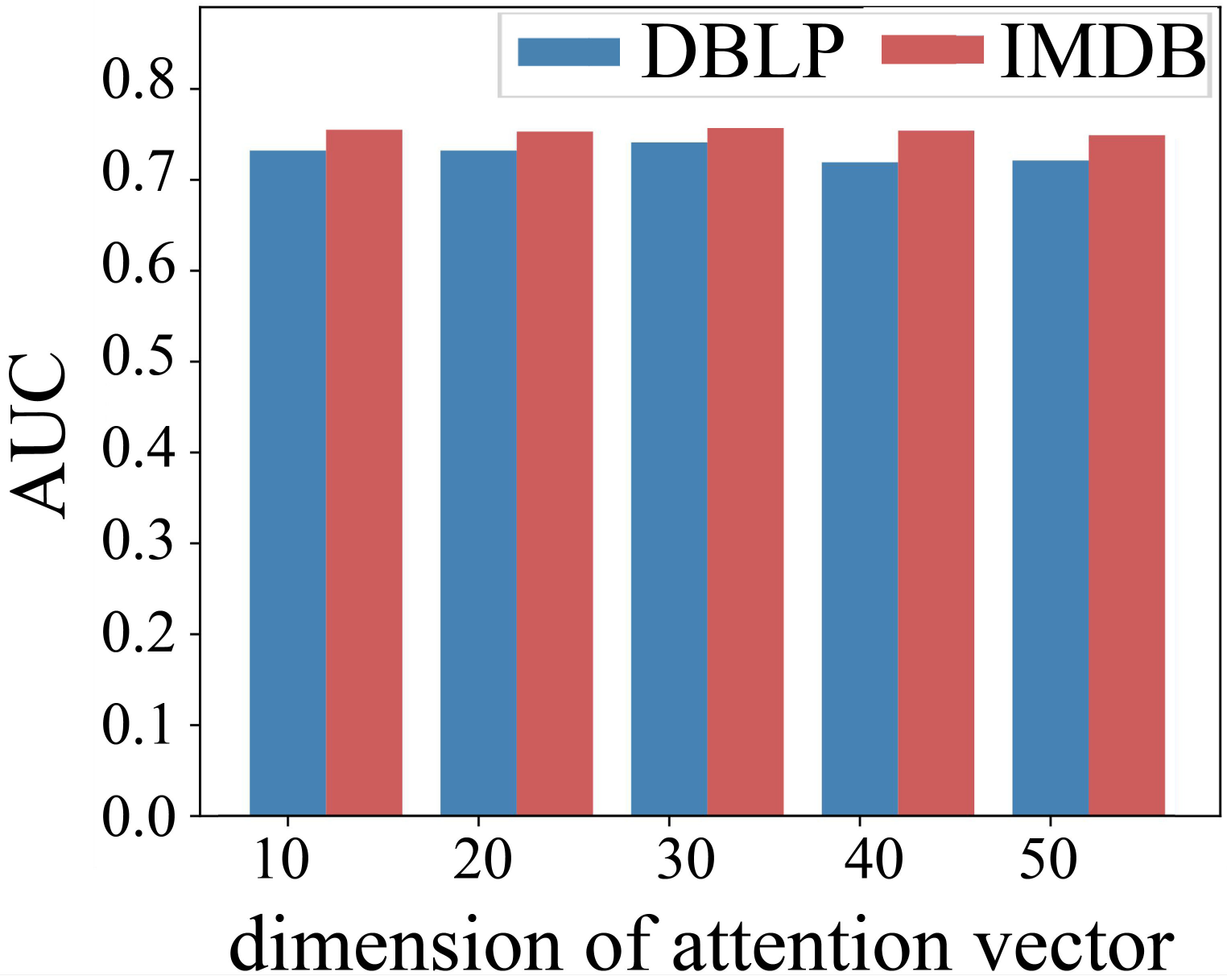}
        \label{Fig.sub.attenauc}
    }
    \caption{the impact of the {dimensions} of {embedding} and attention vector of AUC metric on both DBLP and IMDB datasets.}
    \label{fig:aucmetric}
\end{figure*}

As shown in Figure \ref{Fig.sub.embeddingrecal2} and \ref{Fig.sub.embeddingrecal3}, the dimension of the latent representation has little effect on Recall{@$K$} indicator of AnomMAN. From the statistical data, {on} DBLP dataset, the variation range of Recall@200 and Recall@300 are 0.005 and 0.006 respectively. When {$K$} is 50 or 150, Precision{@$K$} does not change when the dimension of the latent presentation embedding changes. {On} IMDB dataset, when the value of $K$ is chosen as 50, 150, 200 and 300, the variation range of Recall{@$K$} is only 0.08, 0.013, 0.005 and 0.003 respectively.
The impact of the dimension of latent representations on AUC value is shown in Figure \ref{Fig.sub.embeddingauc}. On DBLP dataset, the AUC value varies from 0.737 to 0.745 and the value varies from 0.750 to 0.770 on IMDB dataset.

All these parameter setting values outperform all baselines and 30 is a good choice of the dimension of the latent representations.

\subsubsection{The dimension of attention vector $\mathbf{q}$}

\begin{figure*}
    \centering
    \subfigure[the impact of the dimension of attention vector $\mathbf{q}$ on Recall{@$K$} {on} DBLP dataset]{
        \includegraphics[width=0.45\textwidth]{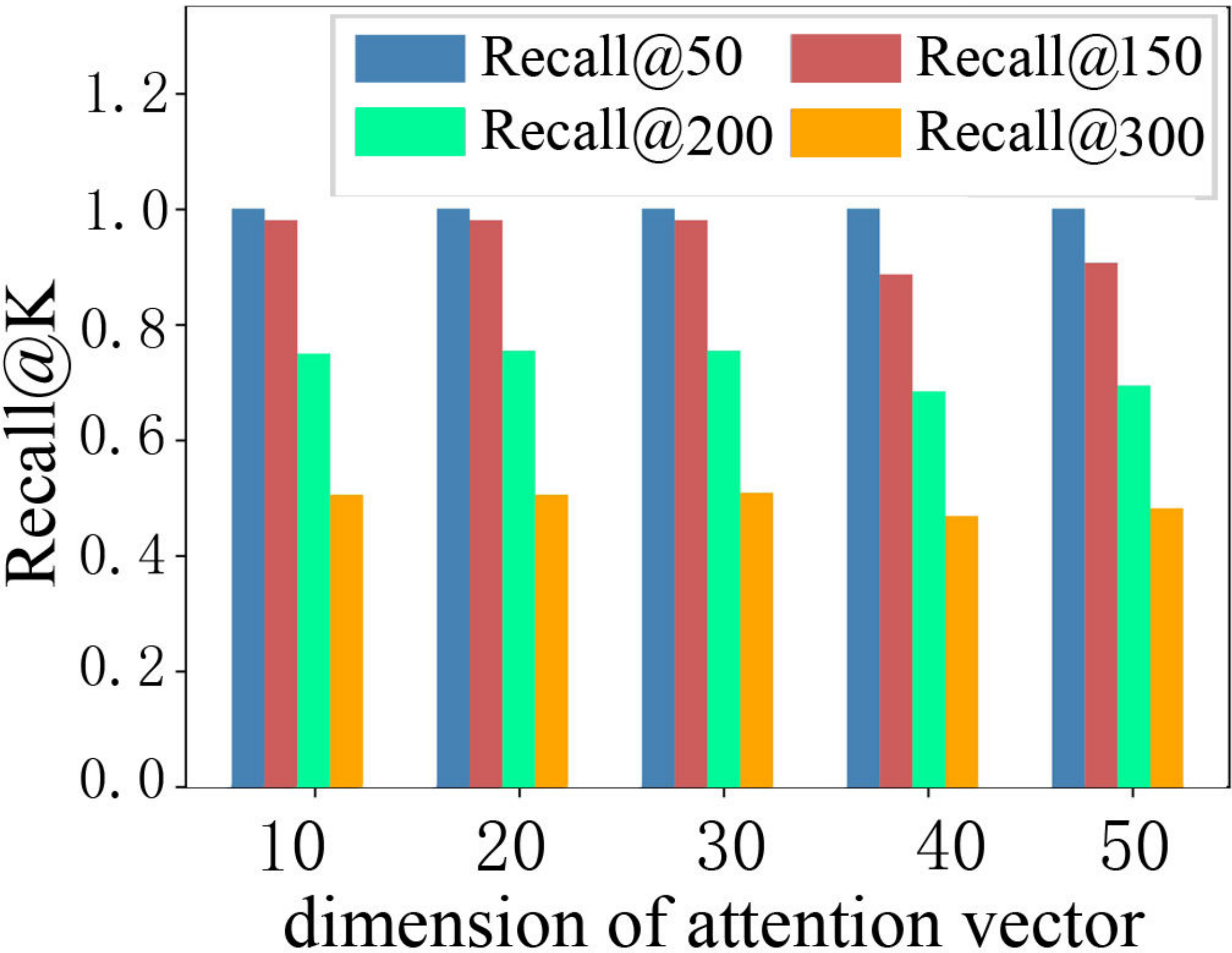}
        \hspace{0mm}
        \label{Fig.sub.attendblp}
    }
    \subfigure[the impact of the dimension of attention vector $\mathbf{q}$ on Recall{@$K$} {on} IMDB dataset]{
        \includegraphics[width=0.45\textwidth]{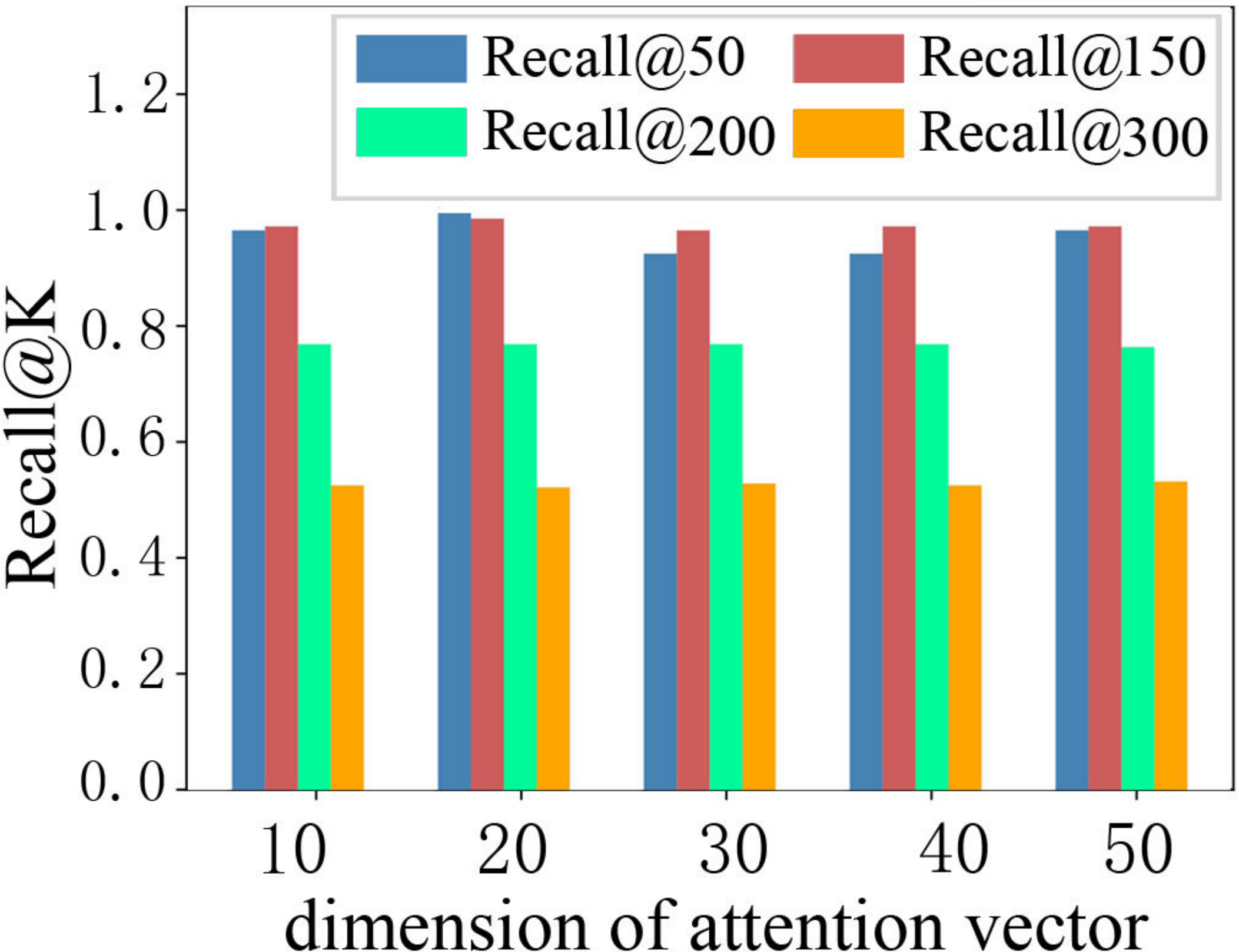}
        \label{Fig.sub.attenimdb}
    }
    \caption{the impact of the dimension of attention vector $\mathbf{q}$ on both DBLP and IMDB datasets.}
    \label{fig:atten}
\end{figure*}

{The} performance of the AnomMAN framework is related to the dimension of the attention vector formulated in Equation \ref{eq:eight}. We chose 5 different choices of dimensions (10, 20, 30, 40, 50) of the latent representation layers on both DBLP and IMDB datasets. The results are shown in {Figure \ref{fig:atten}}.

As shown in Figure \ref{Fig.sub.attendblp} and \ref{Fig.sub.attenimdb}, the dimension of the attention vector $\mathbf{q}$ has little effect on Recall{@$K$} indicator of AnomMAN. From the statistical data, {on} the DBLP dataset, the Recall@50 indicator keeps 1.0 when the dimension of $\mathbf{q}$ varies from 10 to 50. When the dimension size of the attention vector is larger than 30, the Recall{@$K$} indicators drop slightly, but the performance of the model still performs well. {On} the IMDB dataset, when the value of $K$ is chosen as 150, 200, and 300, the variation range of Recall{@$K$} is only 0.02, 0.005, and 0.007 respectively. And the Recall@50 varies from 0.920 to 1.000.
The impact of the dimension of the attention vector $\mathbf{q}$ on the AUC value is shown in Figure \ref{Fig.sub.attenauc}. On the DBLP dataset, the AUC value varies from 0.723 to 0.745 and the value varies from 0.753 to 0.770 on the IMDB dataset.

All these parameter setting values outperform all baselines and 30 is a good choice of the dimension of the attention vector $\mathbf{q}$.

\subsection{Spectral Analysis\label{sec:spec}}

\begin{figure*}
    \centering
    \vspace{-0.8cm} 
    \begin{overpic}[scale=0.56]{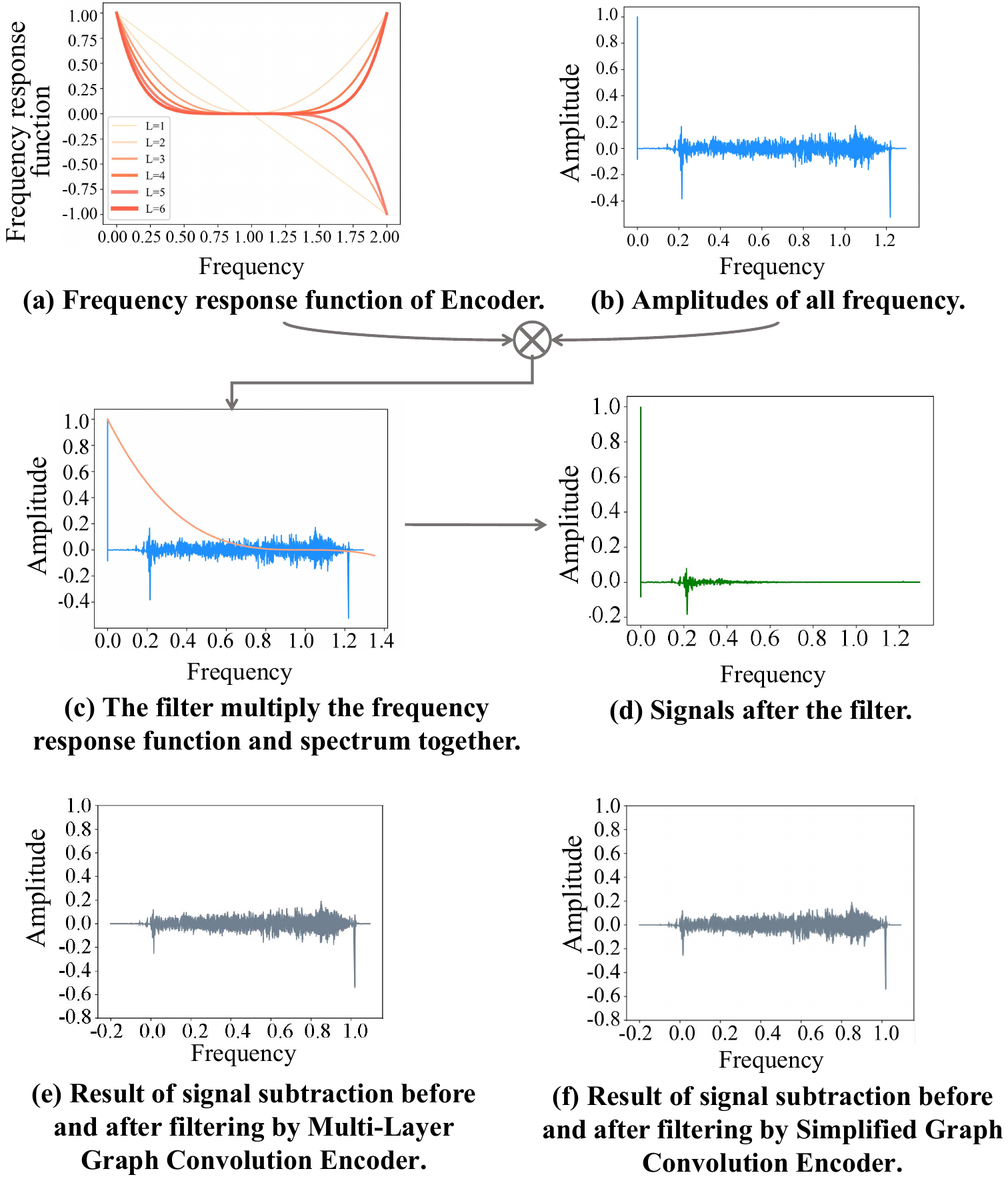}
    \end{overpic}
    \caption{
        The low-pass filter of the multi-view attributed encoder.
    }
    \label{fig:decompose}
\end{figure*}

To verify the performance of a simpler but more powerful low-pass filter, we perform spectral analysis on two datasets. We get that the maximum frequency values of the adjacency matrices $\{A^{(\varphi_k)}\}_{k=1}^{3}$ on the DBLP dataset are 1.400, 1.157, 1.182 respectively and those on IMDB dataset are 1.027, 1.061, 1.293 respectively. All of them are below 1.5. Figure \ref{fig:decompose} shows the performance of the simplified low-pass filter on the IMDB dataset in the co-actor view when $L=3$. In the frequency domain, the filter filters most high-frequency signals and retains most low-frequency signals, thereby realizing the effect of low-pass filtering. It can be seen in Figure~\ref{fig:decompose} that high-frequency signals (between 0.5 to 1.5) are mostly filtered and low-frequency signals (between 0 to 0.5) are mostly reserved. As most anomalous (high-frequency) signals are filtered, instances with a larger error between the inputs and reconstruction representations of AnomMAN are considered to be more anomalous ones.

\subsection{Analysis of Attention Mechanism for information fusion on Multi-view Attributed Networks}

\begin{figure}[hp]
    \centering
    \subfigure[Attention values of different views when the structure anomaly scale of each view is 1:8:1.]{
        \includegraphics[width=0.4\textwidth]{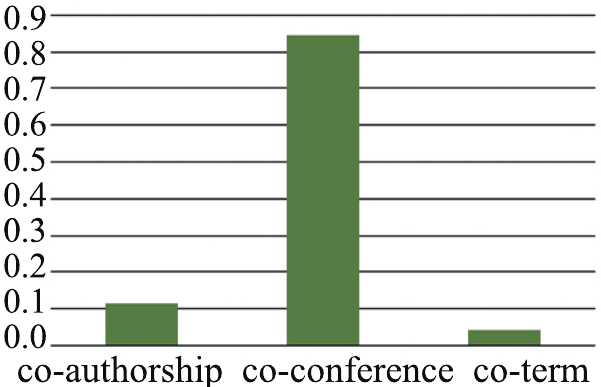}
        \hspace{0mm}
        \label{Fig.sub.att181-dblp}
    }
    \subfigure[Attention values of different views when the structure anomaly scale of each view is 8:1:1.]{
        \includegraphics[width=0.45\textwidth]{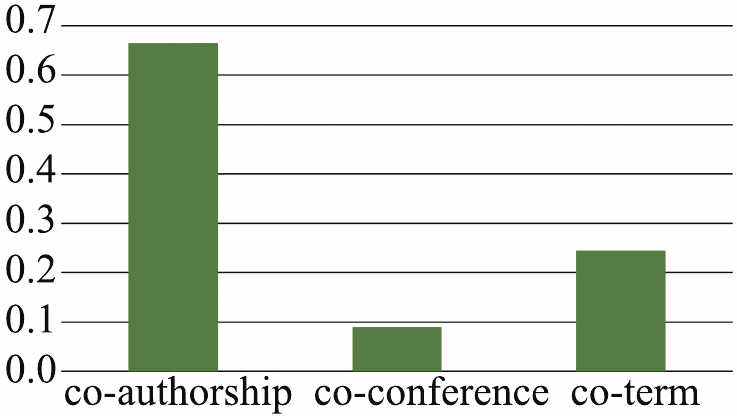}
        \label{Fig.sub.att811-dblp}
    }
    \caption{Attention values of different views {on} DBLP dataset.}
    \label{fig:att-DBLP}
\end{figure}

\begin{figure}[tp]
    \centering
    \subfigure[Attention values of different views when the structure anomaly scale of each view is 1:1:8.]{
        \includegraphics[width=0.4\textwidth]{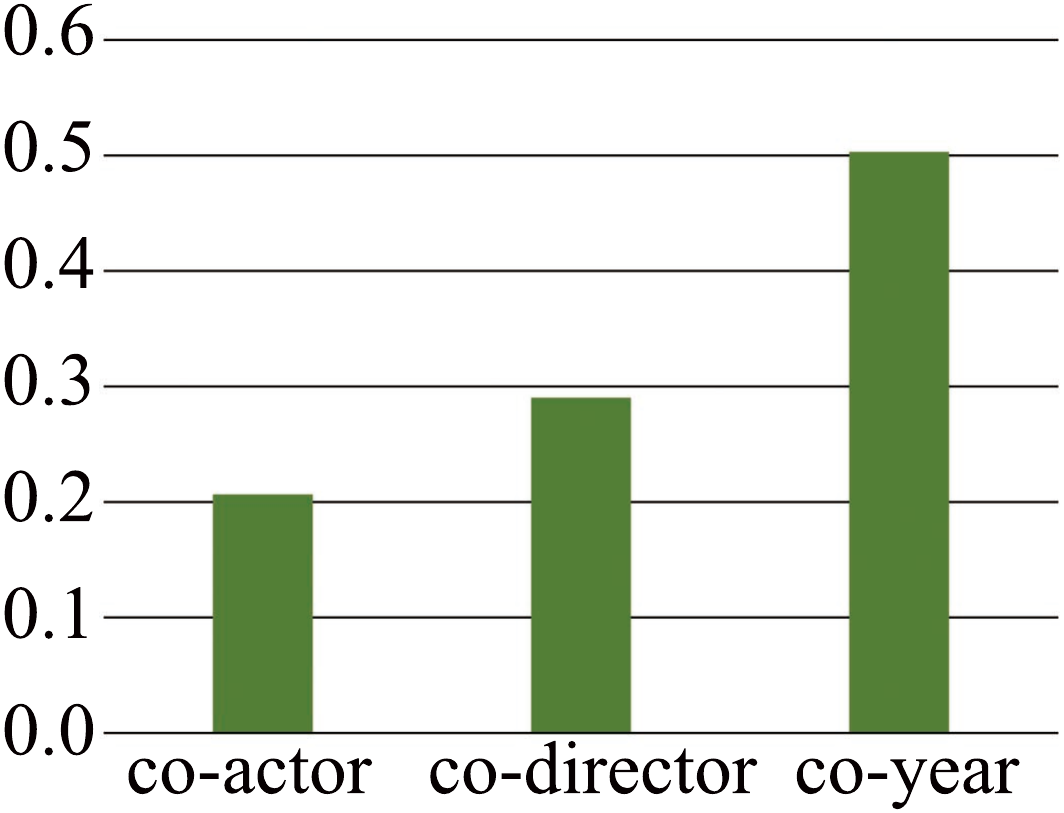}
        \hspace{0mm}
        \label{Fig.sub.att118}
    }
    \subfigure[Attention values of different views when the structure anomaly scale of each view is 8:1:1.]{
        \includegraphics[width=0.4\textwidth]{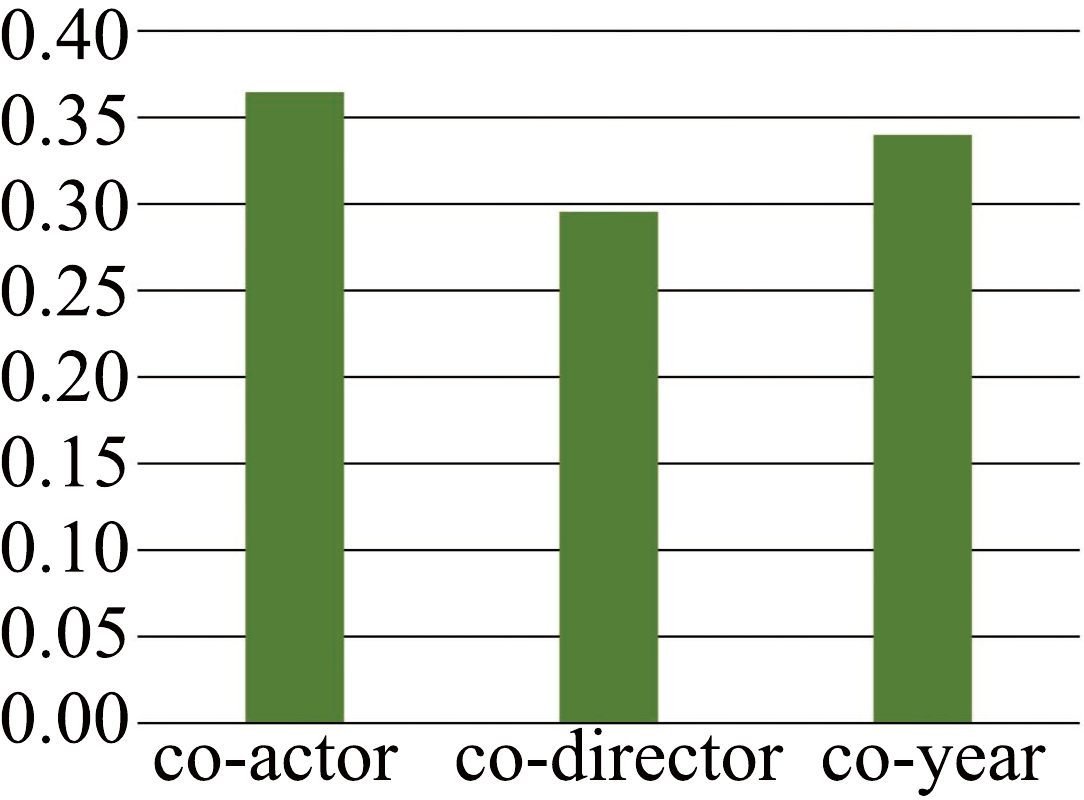}
        \label{Fig.sub.att811}
    }
    \caption{Attention values of different views {on} IMDB dataset.}
    \label{fig:att-IMDB}
\end{figure}

To verify the attention mechanism introduced in Section \ref{sec:attentionfusion}. We analyze the attention value in the view of \textit{co-authorship, co-conference}, and \textit{co-term} {on} DBLP dataset, and \textit{co-actor, co-director}, and \textit{co-year} {on} IMDB dataset. We inject anomalous structures from different views in different scales in DBLP at the ratios of 1:8:1 and 8:1:1 and in IMDB at the ratios of 1:1:8 and 8:1:1. The results of detection are shown in Figures \ref{fig:att-DBLP} and \ref{fig:att-IMDB}. As shown in Figure \ref{Fig.sub.att118}, \textit{co-year} view gets a larger attention value than others when the anomaly ratio is 1:1:8 in the view of \textit{co-actor, co-director}, and \textit{co-year}, respectively. Similarly, as shown in Figure \ref{Fig.sub.att811}, \textit{co-actor} view gets a larger attention value than others when the anomaly ratio is 1:1:8 in the three views, respectively. The phenomenon is the same for DBLP in Figure \ref{fig:att-DBLP}.
\subsection{Visualization}
To compare more intuitively, we conduct the visualization study of node features to verify our motivation.
For anomalous instances, the distances between the input and reconstructed features are large as the high-frequency (anomalous) signals are filtered (analyzed in Section \ref{sec:spec}). Therefore, we analyze the reconstruction error $\Delta \mathbf{x}_i = \mathbf{x}_i - \hat{\mathbf{x}}_i$ between the input and output features for each node. We project the $\Delta \mathbf{\mathbf{x}}_i$ into a 2-dimensional space with t-SNE algorithm \cite{van2008visualizing} to visualize it. We conduct the visualization experiment on both DBLP and IMDB datasets.
\begin{figure}[!h]
    \centering 
    \subfigure[The reconstruction error $\Delta \mathbf{x}_i = \mathbf{x}_i - \hat{\mathbf{x}}_i$ on DBLP dataset.]{
        \includegraphics[width=0.45\textwidth]{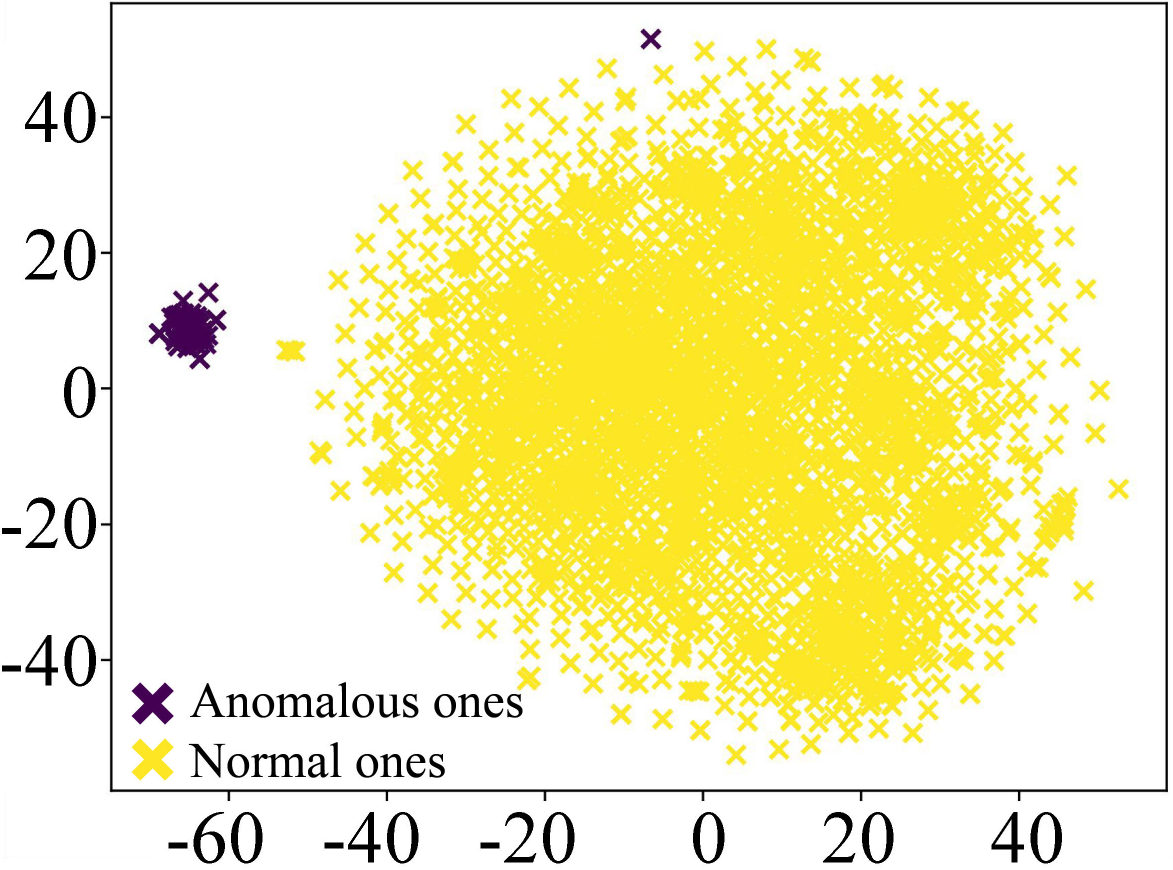}
        \hspace{0mm}
        \label{Fig.sub.tsne-dblp}
    }
    \subfigure[The reconstruction error $\Delta \mathbf{x}_i = \mathbf{x}_i - \hat{\mathbf{x}}_i$ on IMDB dataset.]{
        \includegraphics[width=0.45\textwidth]{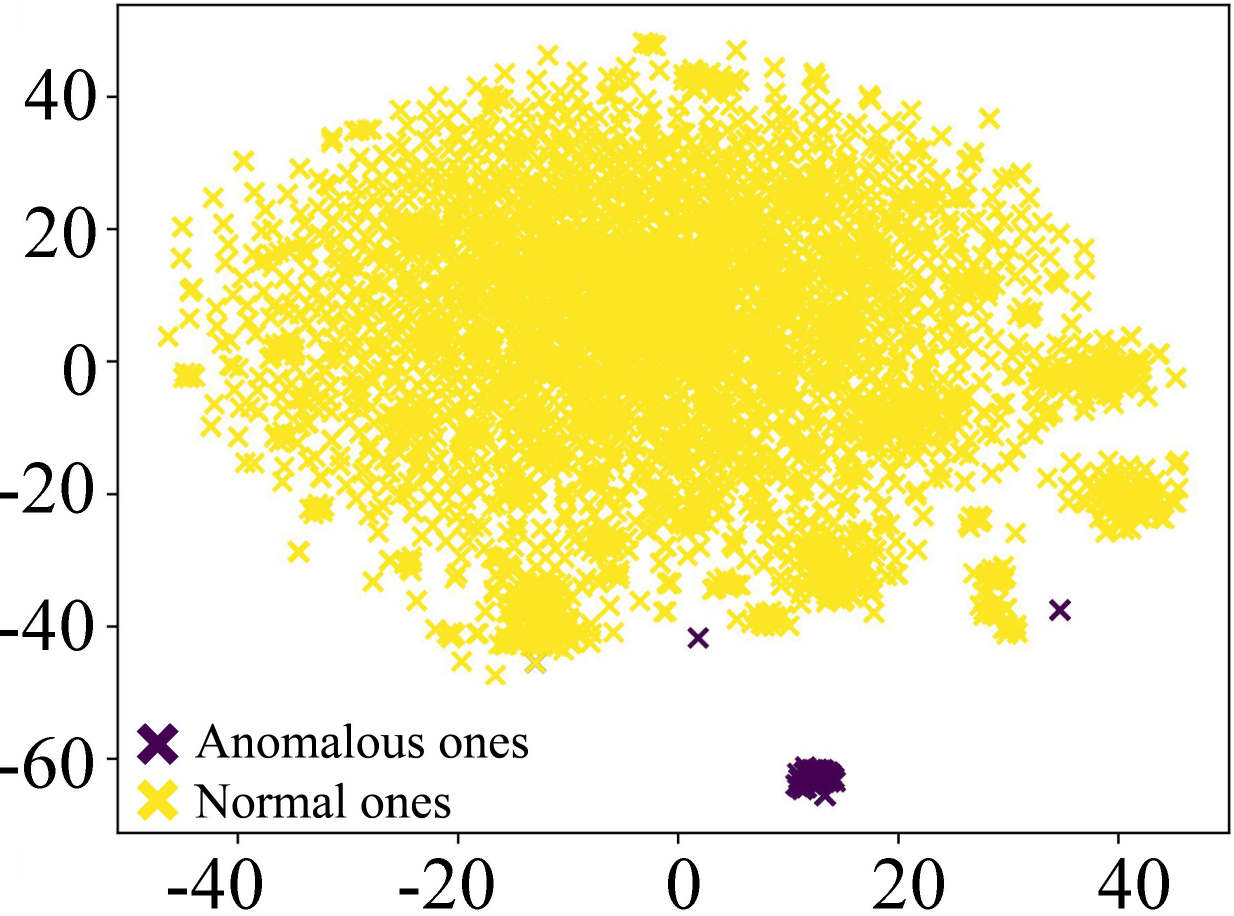}
        \label{Fig.sub.tsne-imdb}
    }
    \caption{Visualization of the reconstruction error $\Delta \mathbf{x}_i = \mathbf{x}_i - \hat{\mathbf{x}}_i$ on DBLP and IMDB datasets. Each '$\times$' notation denotes an instance. }
    \label{fig:my_label}
\end{figure}

As shown in Figure \ref{Fig.sub.tsne-dblp} and \ref{Fig.sub.tsne-imdb}, the reconstruction errors of anomalous ones (purple ones) are much more different from the normal ones (yellow ones). Therefore, those anomalous instances whose high-frequency signals are mostly filtered have larger reconstruction errors and can be detected much easier in the AnomMAN framework.

\section{Conclusion\label{conclusion}}
In this paper, we propose the \textbf{AnomMAN} (Detect \textbf{Anom}aly on \textbf{M}ulti-view \textbf{A}ttributed \textbf{N}etworks) model, which aims to detect anomalies on multi-view attributed networks for the first time. The AnomMAN solves the limitations of the current algorithms in anomaly detection on multi-view attributed networks. AnomMAN uses a graph auto-encoder-based model to overcome the shortcomings of graph convolution as a low-pass filter in anomaly detection tasks. {Furthermore}, it uses a stronger low-pass filter as the multi-view attributed {network} encoder of AnomMAN and shows better performance in this task. The experimental results on two real-world datasets demonstrate that AnomMAN outperforms {both} state-of-the-art models and two variants of our model. Relevant supplementary experiments also fully demonstrate the rationality of our designed model and related research motivations.
In this work, we take advantage of the “\textit{shortcomings}” of graph convolution as a low-pass filter in this anomaly detection task.
However, it is also a possible way to detect anomaly signals by designing a high-pass filter directly. In this work, we try to detect anomalies in an unsupervised fashion. However, it is also possible to detect anomalies with a pre-trained model or some semi-supervised algorithms. In addition, due to {the} large amount of data, there is still a certain distance to apply the anomaly detection algorithm to industrial scenarios. Therefore, our future research will also focus on how to resolve the problem of anomaly detection in very large-scale networks {in} distributed machine learning systems.

\clearpage

\section*{Acknowledgements}
L.H. Chen would like to thank Mr. Wenhao Yang from Nanjing University for providing significant suggestions and support. 
The work was supported by the National Key R\&D Program of China under Grant 2021ZD0201300, the National Research Foundation of Korea(NRF) grant funded by the Korea government(MSIT), (No. 2022R1A2B5B02002456) and by the MSIT(Ministry of Science and ICT), Korea, under the Grand Information Technology Research Center support program(IITP-2023-2020-0-01462) supervised by the IITP(Institute for Information \& communications Technology Planning \& Evaluation).

\clearpage


\begin{thebibliography}{10}

\bibitem{anjaneyulu2019financial}
M~Anjaneyulu and Asst Prof A~Uday KISHORE.
\newblock Financial fraud detection with anomaly feature detection on credit
  card.
\newblock {\em International Journal of Scientific Research \& Engineering},
  5(3):2395--566X, 2019.

\bibitem{bandyopadhyay2019outlier}
Sambaran Bandyopadhyay, N~Lokesh, and M~Narasimha Murty.
\newblock Outlier aware network embedding for attributed networks.
\newblock In {\em Proceedings of the AAAI conference on artificial
  intelligence}, volume~33, pages 12--19, 2019.

\bibitem{bandyopadhyay2020outlier}
Sambaran Bandyopadhyay, Saley~Vishal Vivek, and MN~Murty.
\newblock Outlier resistant unsupervised deep architectures for attributed
  network embedding.
\newblock In {\em Proceedings of the 13th International Conference on Web
  Search and Data Mining}, pages 25--33, 2020.

\bibitem{bo2020structural}
Deyu Bo, Xiao Wang, Chuan Shi, Meiqi Zhu, Emiao Lu, and Peng Cui.
\newblock Structural deep clustering network.
\newblock In {\em Proceedings of The Web Conference 2020}, pages 1400--1410,
  2020.

\bibitem{breunig2000lof}
Markus~M Breunig, Hans-Peter Kriegel, Raymond~T Ng, and J{\"o}rg Sander.
\newblock Lof: identifying density-based local outliers.
\newblock In {\em Proceedings of the 2000 ACM SIGMOD international conference
  on Management of data}, pages 93--104, 2000.

\bibitem{bruna2013spectral}
Joan Bruna, Wojciech Zaremba, Arthur Szlam, and Yann LeCun.
\newblock Spectral networks and locally connected networks on graphs.
\newblock {\em arXiv preprint arXiv:1312.6203}, 2013.

\bibitem{chen2020graph}
Fenxiao Chen, Yun-Cheng Wang, Bin Wang, and C-C~Jay Kuo.
\newblock Graph representation learning: A survey.
\newblock {\em APSIPA Transactions on Signal and Information Processing}, 9,
  2020.

\bibitem{chen2022learning}
Yihao Chen, Xin Tang, Xianbiao Qi, Chun-Guang Li, and Rong Xiao.
\newblock Learning graph normalization for graph neural networks.
\newblock {\em Neurocomputing}, 493:613--625, 2022.

\bibitem{WANG2021700}
{Chen Wang and Xiaojun Chen and Bingkun Chen and Feiping Nie and Bo Wang and
  Zhong Ming}.
\newblock {Learning unsupervised node representation from multi-view network}.
\newblock {\em {Information Sciences}}, {579}:{700--716}, {2021}.

\bibitem{defferrard2016convolutional}
Micha{\"e}l Defferrard, Xavier Bresson, and Pierre Vandergheynst.
\newblock Convolutional neural networks on graphs with fast localized spectral
  filtering.
\newblock {\em Advances in neural information processing systems},
  29:3844--3852, 2016.

\bibitem{ding2019deep}
Kaize Ding, Jundong Li, Rohit Bhanushali, and Huan Liu.
\newblock Deep anomaly detection on attributed networks.
\newblock In {\em Proceedings of the 2019 SIAM International Conference on Data
  Mining}, pages 594--602. SIAM, 2019.

\bibitem{du2022graph}
{Du, Xusheng and Yu, Jiong and Chu, Zheng and Jin, Lina and Chen, Jiaying}.
\newblock {Graph autoencoder-based unsupervised outlier detection}.
\newblock {\em {Information Sciences}}, {608}:{532--550}, {2022}.

\bibitem{Eswaran2018spotlight}
Dhivya Eswaran, Christos Faloutsos, Sudipto Guha, and Nina Mishra.
\newblock Spotlight: Detecting anomalies in streaming graphs.
\newblock In {\em Proceedings of the 24th ACM SIGKDD International Conference
  on Knowledge Discovery \& Data Mining}, pages 1378--1386, 2018.

\bibitem{fan2020one2multi}
Shaohua Fan, Xiao Wang, Chuan Shi, Emiao Lu, Ken Lin, and Bai Wang.
\newblock One2multi graph autoencoder for multi-view graph clustering.
\newblock In {\em Proceedings of The Web Conference 2020}, pages 3070--3076,
  2020.

\bibitem{WU2022142}
{Fei Wu and Xiao-Yuan Jing and Pengfei Wei and Chao Lan and Yimu Ji and
  Guo-Ping Jiang and Qinghua Huang}.
\newblock {Semi-supervised multi-view graph convolutional networks with
  application to webpage classification}.
\newblock {\em {Information Sciences}}, {591}:{142--154}, {2022}.

\bibitem{hamilton2017inductive}
Will Hamilton, Zhitao Ying, and Jure Leskovec.
\newblock Inductive representation learning on large graphs.
\newblock In {\em Advances in neural information processing systems}, pages
  1024--1034, 2017.

\bibitem{hooi2016fraudar}
Bryan Hooi, Hyun~Ah Song, Alex Beutel, Neil Shah, Kijung Shin, and Christos
  Faloutsos.
\newblock Fraudar: Bounding graph fraud in the face of camouflage.
\newblock In {\em Proceedings of the 22nd ACM SIGKDD International Conference
  on Knowledge Discovery and Data Mining}, pages 895--904, 2016.

\bibitem{huang2021hybrid}
Ling Huang, Ye~Zhu, Yuefang Gao, Tuo Liu, Chao Chang, Caixing Liu, Yong Tang,
  and Chang-Dong Wang.
\newblock Hybrid-order anomaly detection on attributed networks.
\newblock {\em IEEE Transactions on Knowledge and Data Engineering}, 2021.

\bibitem{huang2020biane}
Wentao Huang, Yuchen Li, Yuan Fang, Ju~Fan, and Hongxia Yang.
\newblock Biane: Bipartite attributed network embedding.
\newblock In {\em Proceedings of the 43rd international ACM SIGIR conference on
  research and development in information retrieval}, pages 149--158, 2020.

\bibitem{huang2018exploring}
Xiao Huang, Qingquan Song, Jundong Li, and Xia Hu.
\newblock Exploring expert cognition for attributed network embedding.
\newblock In {\em Proceedings of the Eleventh ACM International Conference on
  Web Search and Data Mining}, pages 270--278, 2018.

\bibitem{kingma2014adam}
Diederik~P Kingma and Jimmy Ba.
\newblock Adam: A method for stochastic optimization.
\newblock {\em arXiv preprint arXiv:1412.6980}, 2014.

\bibitem{kipf2016semi}
Thomas~N Kipf and Max Welling.
\newblock Semi-supervised classification with graph convolutional networks.
\newblock {\em arXiv preprint arXiv:1609.02907}, 2016.

\bibitem{li2022deep}
He~Li, Xuejiao Li, Liangcai Su, Duo Jin, Jianbin Huang, and Deshuang Huang.
\newblock Deep spatio-temporal adaptive 3d convolutional neural networks for
  traffic flow prediction.
\newblock {\em ACM Transactions on Intelligent Systems and Technology (TIST)},
  13(2):1--21, 2022.

\bibitem{li2021detectornet}
He~Li, Shiyu Zhang, Xuejiao Li, Liangcai Su, Hongjie Huang, Duo Jin, Linghao
  Chen, Jianbin Huang, and Jaesoo Yoo.
\newblock Detectornet: Transformer-enhanced spatial temporal graph neural
  network for traffic prediction.
\newblock In {\em Proceedings of the 29th International Conference on Advances
  in Geographic Information Systems}, pages 133--136, 2021.

\bibitem{li2020graphsanet}
He~Li, Shiyu Zhang, Liangcai Su, Hongjie Huang, Duo Jin, and Xuejiao Li.
\newblock Graphsanet: a graph neural network and self attention based approach
  for spatial temporal prediction in sensor network.
\newblock In {\em 2020 IEEE International Conference on Big Data (Big Data)},
  pages 5756--5758. IEEE, 2020.

\bibitem{li2017radar}
Jundong Li, Harsh Dani, Xia Hu, and Huan Liu.
\newblock Radar: Residual analysis for anomaly detection in attributed
  networks.
\newblock In {\em IJCAI}, pages 2152--2158, 2017.

\bibitem{li2019trust}
Xiaolin Li, Yuan Zhuang, Yanjie Fu, and Xiangdong He.
\newblock A trust-aware random walk model for return propensity estimation and
  consumer anomaly scoring in online shopping.
\newblock {\em Science China Information Sciences}, 62(5):1--17, 2019.

\bibitem{li2019specae}
Yuening Li, Xiao Huang, Jundong Li, Mengnan Du, and Na~Zou.
\newblock Specae: Spectral autoencoder for anomaly detection in attributed
  networks.
\newblock In {\em Proceedings of the 28th ACM international conference on
  information and knowledge management}, pages 2233--2236, 2019.

\bibitem{li2020copula}
Zheng Li, Yue Zhao, N~Botta, C~Ionescu, and X~COPOD Hu.
\newblock Copula-based outlier detection.
\newblock In {\em Proceedings of the 2020 IEEE International Conference on Data
  Mining (ICDM), Sorrento, Italy}, pages 17--20, 2020.

\bibitem{liu2021anomaly}
Yixin Liu, Zhao Li, Shirui Pan, Chen Gong, Chuan Zhou, and George Karypis.
\newblock Anomaly detection on attributed networks via contrastive
  self-supervised learning.
\newblock {\em IEEE transactions on neural networks and learning systems},
  33(6):2378--2392, 2021.

\bibitem{liu2021motif}
Zhijun Liu, Chao Huang, Yanwei Yu, and Junyu Dong.
\newblock Motif-preserving dynamic attributed network embedding.
\newblock In {\em Proceedings of the Web Conference 2021}, pages 1629--1638,
  2021.

\bibitem{ou2016asymmetric}
Mingdong Ou, Peng Cui, Jian Pei, Ziwei Zhang, and Wenwu Zhu.
\newblock Asymmetric transitivity preserving graph embedding.
\newblock In {\em Proceedings of the 22nd ACM SIGKDD international conference
  on Knowledge discovery and data mining}, pages 1105--1114, 2016.

\bibitem{peng2018anomalous}
Zhen Peng, Minnan Luo, Jundong Li, Huan Liu, and Qinghua Zheng.
\newblock Anomalous: A joint modeling approach for anomaly detection on
  attributed networks.
\newblock In {\em IJCAI}, pages 3513--3519, 2018.

\bibitem{peng2020deep}
Zhen Peng, Minnan Luo, Jundong Li, Luguo Xue, and Qinghua Zheng.
\newblock A deep multi-view framework for anomaly detection on attributed
  networks.
\newblock {\em IEEE Transactions on Knowledge and Data Engineering}, 2020.

\bibitem{perozzi2016scalable}
Bryan Perozzi and Leman Akoglu.
\newblock Scalable anomaly ranking of attributed neighborhoods.
\newblock In {\em Proceedings of the 2016 SIAM International Conference on Data
  Mining}, pages 207--215. SIAM, 2016.

\bibitem{ren2019time}
Hansheng Ren, Bixiong Xu, Yujing Wang, Chao Yi, Congrui Huang, Xiaoyu Kou, Tony
  Xing, Mao Yang, Jie Tong, and Qi~Zhang.
\newblock Time-series anomaly detection service at microsoft.
\newblock In {\em Proceedings of the 25th ACM SIGKDD International Conference
  on Knowledge Discovery \& Data Mining}, pages 3009--3017, 2019.

\bibitem{rumelhart1986learning}
David~E Rumelhart, Geoffrey~E Hinton, and Ronald~J Williams.
\newblock Learning representations by back-propagating errors.
\newblock {\em nature}, 323(6088):533--536, 1986.

\bibitem{scarselli2008graph}
Franco Scarselli, Marco Gori, Ah~Chung Tsoi, Markus Hagenbuchner, and Gabriele
  Monfardini.
\newblock The graph neural network model.
\newblock {\em IEEE transactions on neural networks}, 20(1):61--80, 2008.

\bibitem{shang2014modbus}
Wen-li SHANG, Sheng-shan ZHANG, Ming WAN, and Peng ZENG.
\newblock Modbus/tcp communication anomaly detection algorithm based on
  pso-svm.
\newblock {\em ACTA ELECTONICA SINICA}, 42(11):2314, 2014.

\bibitem{van2008visualizing}
Laurens Van~der Maaten and Geoffrey Hinton.
\newblock Visualizing data using t-sne.
\newblock {\em Journal of machine learning research}, 9(11), 2008.

\bibitem{vaswani2017attention}
Ashish Vaswani, Noam Shazeer, Niki Parmar, Jakob Uszkoreit, Llion Jones,
  Aidan~N Gomez, {\L}ukasz Kaiser, and Illia Polosukhin.
\newblock Attention is all you need.
\newblock In {\em Advances in neural information processing systems}, pages
  5998--6008, 2017.

\bibitem{velivckovic2017graph}
Petar Veli{\v{c}}kovi{\'c}, Guillem Cucurull, Arantxa Casanova, Adriana Romero,
  Pietro Lio, and Yoshua Bengio.
\newblock Graph attention networks.
\newblock {\em arXiv preprint arXiv:1710.10903}, 2017.

\bibitem{wang2017community}
Xiao Wang, Peng Cui, Jing Wang, Jian Pei, Wenwu Zhu, and Shiqiang Yang.
\newblock Community preserving network embedding.
\newblock In {\em Proceedings of the AAAI Conference on Artificial
  Intelligence}, volume~31, 2017.

\bibitem{wu2019simplifying}
Felix Wu, Amauri Souza, Tianyi Zhang, Christopher Fifty, Tao Yu, and Kilian
  Weinberger.
\newblock Simplifying graph convolutional networks.
\newblock In {\em International conference on machine learning}, pages
  6861--6871. PMLR, 2019.

\bibitem{wu2020comprehensive}
Zonghan Wu, Shirui Pan, Fengwen Chen, Guodong Long, Chengqi Zhang, and S~Yu
  Philip.
\newblock A comprehensive survey on graph neural networks.
\newblock {\em IEEE transactions on neural networks and learning systems},
  32(1):4--24, 2020.

\bibitem{xia2023moderate}
Xiaobo Xia, Jiale Liu, Jun Yu, Xu~Shen, Bo~Han, and Tongliang Liu.
\newblock Moderate coreset: A universal method of data selection for real-world
  data-efficient deep learning.
\newblock In {\em The Eleventh International Conference on Learning
  Representations}, 2023.

\bibitem{xia2022pluralistic}
Xiaobo Xia, Wenhao Yang, Jie Ren, Yewen Li, Yibing Zhan, Bo~Han, and Tongliang
  Liu.
\newblock Pluralistic image completion with gaussian mixture models.
\newblock In {\em NeurIPS}, 2022.

\bibitem{xu2019graph}
Bingbing Xu, Huawei Shen, Qi~Cao, Yunqi Qiu, and Xueqi Cheng.
\newblock Graph wavelet neural network.
\newblock {\em arXiv preprint arXiv:1904.07785}, 2019.

\bibitem{yu2018netwalk}
Wenchao Yu, Wei Cheng, Charu~C Aggarwal, Kai Zhang, Haifeng Chen, and Wei Wang.
\newblock Netwalk: A flexible deep embedding approach for anomaly detection in
  dynamic networks.
\newblock In {\em Proceedings of the 24th ACM SIGKDD International Conference
  on Knowledge Discovery \& Data Mining}, pages 2672--2681, 2018.

\bibitem{zhang2022reconstruction}
Jiaqiang Zhang, Senzhang Wang, and Songcan Chen.
\newblock Reconstruction enhanced multi-view contrastive learning for anomaly
  detection on attributed networks.
\newblock {\em arXiv preprint arXiv:2205.04816}, 2022.

\bibitem{zheng2021generative}
Yu~Zheng, Ming Jin, Yixin Liu, Lianhua Chi, Khoa~T. Phan, and Yi-Ping~Phoebe
  Chen.
\newblock Generative and contrastive self-supervised learning for graph anomaly
  detection.
\newblock {\em IEEE Transactions on Knowledge and Data Engineering}, 2021.

\bibitem{zhou2018graph}
Jie Zhou, Ganqu Cui, Zhengyan Zhang, Cheng Yang, Zhiyuan Liu, Lifeng Wang,
  Changcheng Li, and Maosong Sun.
\newblock Graph neural networks: A review of methods and applications.
\newblock {\em arXiv preprint arXiv:1812.08434}, 2018.

\bibitem{zhu2022bars}
Jieming Zhu, Quanyu Dai, Liangcai Su, Rong Ma, Jinyang Liu, Guohao Cai,
  Xi~Xiao, and Rui Zhang.
\newblock Bars: Towards open benchmarking for recommender systems.
\newblock In {\em Proceedings of the 45th International ACM SIGIR Conference on
  Research and Development in Information Retrieval}, pages 2912--2923, 2022.

\end{thebibliography}
\end{document}